\documentclass[authoryear,3p,11pt]{elsarticle}
\usepackage{amsmath,amssymb,mathtools,mathrsfs}

\usepackage{ulem}
\usepackage{cancel}
\usepackage{verbatim}
\usepackage{float}

\usepackage{array}

\usepackage[pdftex, pdftitle={Article}, pdfauthor={Author}]{hyperref} % For hyperlinks in the PDF
\usepackage[usenames,dvipsnames]{xcolor}
\usepackage{booktabs}
\usepackage{multirow}

\newcommand{\upd}{\mathrm{d}}
\newcommand{\dL}{\Delta L}
\newcommand{\lp}{\ell_p}
\newcommand{\etastab}{\eta_{\mathrm{stab}}}
\newcommand{\crit}{\mathrm{crit}}
\newcommand{\lambdad}{\lambda_d}
\newcommand{\lambdas}{\lambda_s}

\journal{Elsevier}

\date{October 19, 2020}

\begin{document}

\begin{frontmatter}

%\title{Frustrating frustration in snap-induced morphing}

\title{Snap-induced morphing: From a single bistable shell to    the  origin of shape bifurcation in interacting shells}


\author{Mingchao Liu$^{\dag}$, Lucie Domino$^{\dag}$, Iris Dupont de Dinechin$^{\dag}$, Matteo Taffetani$^{\ddag}$ and Dominic Vella$^{\dag}$}

\address{$^{\dag}$\:Mathematical Institute, University of Oxford, Woodstock Rd, Oxford, OX2 6GG, UK\\
$^{\ddag}$\:Department of Engineering Mathematics, University of Bristol, Bristol, BS8 1TW, UK}

\begin{abstract}
The bistability of embedded elements provides a natural route through which to introduce reprogrammability to elastic meta-materials. One example of this is the soft morphable sheet, in which bistable elements that can be snapped up or down, are embedded within a soft sheet. The state of the sheet can then be programmed by snapping particular elements up or down, resulting in different global shapes. However, attempts to leverage this programmability have been limited by the tendency for the deformations induced by multiple elastic elements to cause large global shape bifurcations. We study the root cause of this bifurcation in the soft morphable sheet by developing a detailed understanding of the behaviour of a single bistable element attached to a flat ‘skirt’ region. We study the geometrical limitations on the bistability of this single element, and show that the structure of its deformation can be understood using a boundary layer analysis. Moreover, by studying the compressive strains that a single bistable element induces in the surrounding skirt we show that the shape bifurcation in the soft morphable sheet can be delayed by an appropriate design of the lattice on which bistable elements are placed.
\end{abstract}


\end{frontmatter}

\section{Introduction}

The ability to change shape is as important to an emerging class of engineering applications as it is to biological organisms: just as animals and plants morph in response to external stimuli, soft robots must be able to change shape to adapt to different environments and complete different tasks \cite[][]{Alapan2020,Shah2021,liu2021}. In both cases, our understanding of different artificial mechanisms through which this shape change can be achieved has exploded in recent years with examples including pneumatic inflation \cite[][]{Pikul2017,siefert2019}, multi-material 4D printing \cite[][]{boley2019}, magneto-responsive elastomers \cite[][]{zhang2021} and 3D-printed composites \cite[][]{kim2018}, designed director fields within liquid crystal elastomers \cite[][]{aharoni2018}, designed cuts in planar sheets \cite[][]{celli2018,liu2020}, as well as swelling of patterned hydrogels \cite[]{wang2017}. 

In each of the above artificial examples, techniques have been developed that allow a particular three-dimensional target shape to be achieved by the actuation of an initially flat sheet. However, these techniques in general are only capable of generating one, or perhaps two, designed shapes. Nature, however, is able to adapt form repeatedly and in a variety of ways. To achieve something similar in artificial systems requires a means of reprogramming the shape.

There are two key hurdles to achieve this reprogrammability in practice: firstly, a means of reversibly actuating different elements of an object between two different states \cite[][]{faber2020,Alapan2020}. Secondly, interactions between neighbouring elements may lead to unwanted global deformation mode that make it impossible to reach a well-controlled state \cite[][]{moessner2006,gilbert2016,siefert2021}: the system is then generically stuck in a local minimum, with many similar states `nearby' --- what may be termed `soft modes'.

In this paper, we study a soft morphable sheet, illustrating another system in which shape can be changed by controlling the state of individual elements; we show how understanding the behaviour of these individual elements may yield new understanding, allowing the appearance of soft modes to be postponed.

\begin{figure}[ht]
    \centering
    \vspace{0.5cm}
    \includegraphics[width=0.55\linewidth]{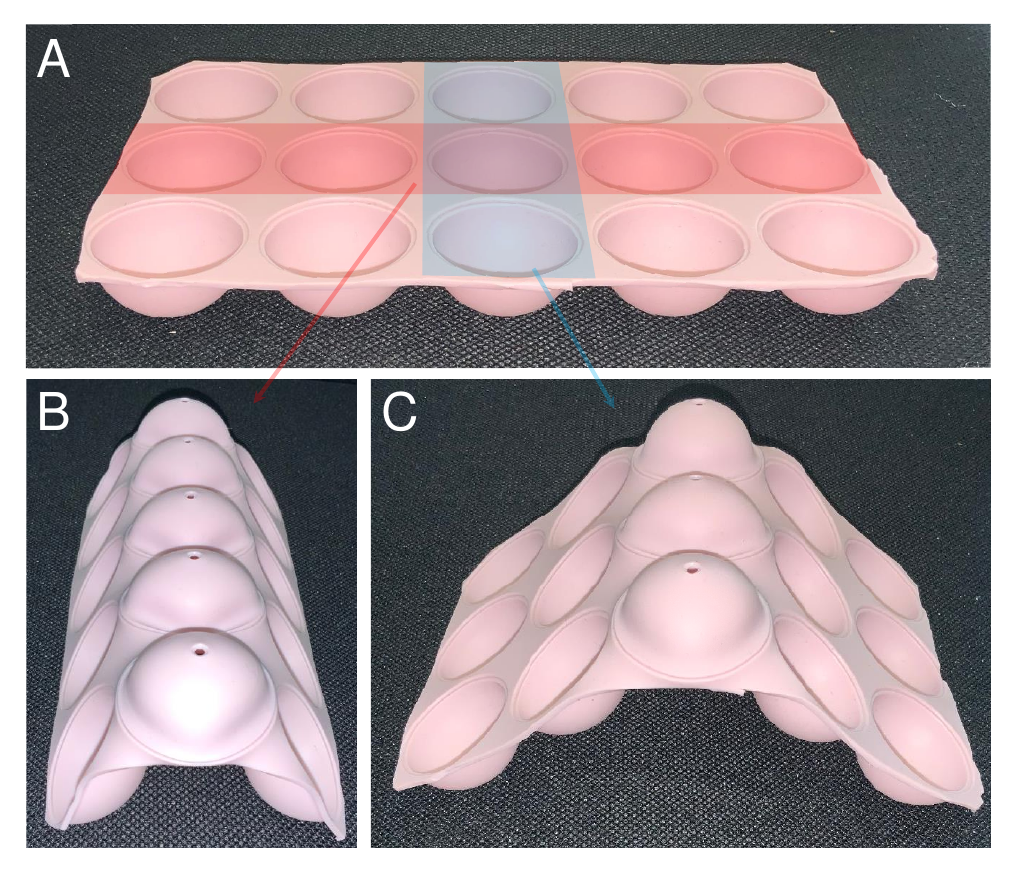}
    \caption{A silicone chocolate mould is an everyday instance of a `Soft Morphing Sheet': by popping individual hemispherical elements, the mould morphs from its initial flat state (shown in (A)) to different stable curved states ((B) and (C)). In (B) the sheet curves around its long axis (highlighted in red in panel A) in response to elements being popped along this axis. In (C) the sheet curves about its short axis (highlighted in blue in panel A) in response to elements along the short axis being popped. (The diameter of each spherical cap is $3.5\mathrm{~cm}$.) }
    \label{fig:ChocolateMould}
\end{figure}

To achieve a fundamental element with two different stable (activated and natural, or de-activated) states, elastic snap-through is a natural candidate: curved elastic structures such as arches and shells often have two stable states and can be switched between these two states by the application of a suitable external load \cite[][]{holmes2007,Taffetani2018,liu2021delayed}. It is therefore somewhat natural to consider a simple elastic sheet with multiple bistable elements embedded at various locations within it. Conveniently, a similar system can be found in the silicone moulds used for moulding chocolate (see Fig.~\ref{fig:ChocolateMould}). Popping individual hemispherical dimples (the moulds) from their natural state to inverted state induces deformations in the neighbouring sheet that is reminiscent of the deformation induced by a localized dilation \cite[][]{Oshri2019,Plummer2020,Oshri2020,hanakata2022}. However, as more snapping elements are activated, we see that the shape of the whole sheet changes globally from the initially flat state (Fig.~\ref{fig:ChocolateMould}A) to different stable curved states (Fig.~\ref{fig:ChocolateMould}B and C), depending on which dimples one chooses to pop. In this sense, the silicone mould represents a ``\emph{Soft Morphing Sheet}''. A rigid version of this idea was first presented by Seffen \cite[][]{seffen2006, seffen2007}, but the use of an elastomer here allows deeper shells to pop reversibly (without plastic deformation), thereby creating larger local deformations and, as a result, greater global shape changes. We seek to understand this deformation, and so shall not focus on the means of switching the individual elements between different states, merely noting that this could be achieved using pneumatic pressure \cite[][]{gorissen2020} or externally applied magnetic fields \cite[][]{Chen2021}, for example.

\begin{figure}%[tbhp]
    \centering
    \vspace{0.5cm}
    \includegraphics[width=0.6\linewidth]{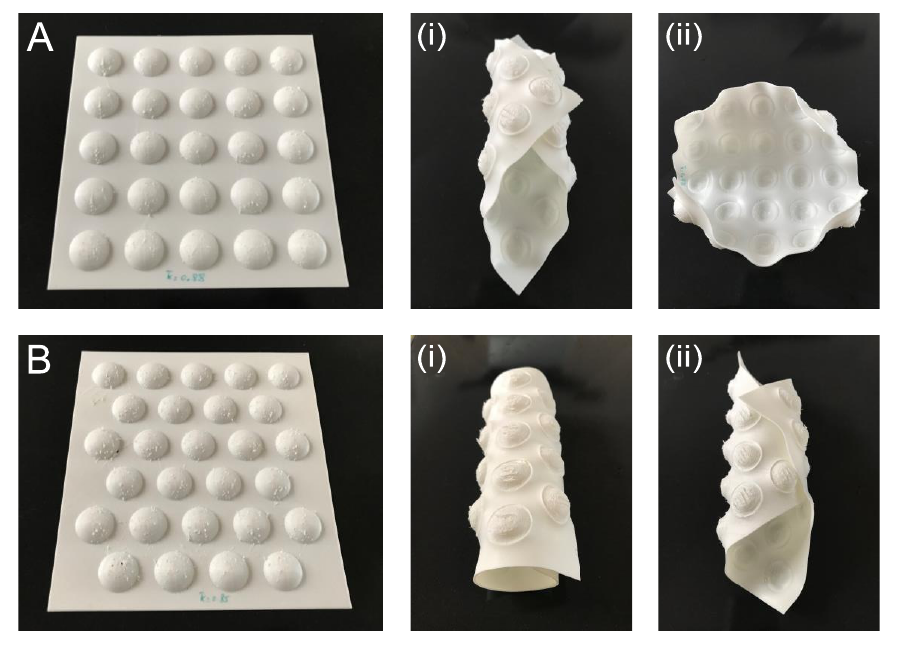}
    \caption{The frustration of inverted dimples placed on a square lattice (top) and triangular lattice (bottom) leads to different global modes of deformation that can be easily shifted by hand (`soft modes'). (A) The natural (undeformed) state of the sheet with natural dimples on a square lattice (with total size of 159 $\times$ 159 $\mathrm{~mm}$) leads to at least two deformation shapes when the dimples are inverted, (A-i) and (A-ii). (B)  The natural (undeformed) state of the sheet with natural dimples on a triangular lattice (169 $\times$ 157 $\mathrm{~mm}$) leads to at least two deformation shapes when the dimples are inverted, (B-i) and (B-ii). In each case these sheets are 3D printed as described in  \ref{methods_exp}.}
    \label{fig:Frustration}
\end{figure}

A similar system has been studied recently \cite[][]{udani2022taming}. This work showed that the system becomes frustrated with many co-existing states with similar energies: geometrical frustration of neighbouring elements leads to degeneracy of the global shape, which are sometimes called `soft modes' of deformation \cite[][]{moessner2006}. The system can be switched between neighbouring frustrated states by gently teasing the system (e.g.~by hand); however, this frustration cannot be eliminated. Examples of the different global shapes that can be achieved as a result of this frustration for dimples on square and  triangular lattices are shown in Fig.~\ref{fig:Frustration}. In the application to shape change,  however, the aim is to control the state of the individual elements to be buckled (or not), and hence to understand how the macroscopic behaviour (i.e.~global shape) emerges from the microscopic behaviour (i.e.~local deformation). In this paper, therefore, we seek to understand the buckling instability that leads to shape bifurcation, as well as illustrating one strategy to delay its onset. 

The paper is organized as follows. We start by analysing the individual snapping shell as a single element in Section \ref{sec:SingleElement}. Here, we make use of shallow shell theory (presented in Section \ref{sec:ShallowShellFormulation}) to study first  the conditions for bistability (Section \ref{sec:Bistability}). To understand the inverted state further we then present a boundary layer analysis of  the deformed shape of a single element (Section \ref{sec:DeformedElementShape}); in each case, we compare our theoretical results with those from both Finite Element Method (FEM) simulations and experiments. The boundary layer analysis of Section \ref{sec:DeformedElementShape} lends insight into the root cause for the shape bifurcation that is observed and results in frustration. We therefore study, in Section \ref{sec:ManyElements}, the effect of lattice design on shape bifurcation and hence on frustration, showing that shape bifurcation can be delayed by placing elements on a hexagonal lattice, rather than a triangular lattice. Finally, in Section \ref{sec:DiscussionandConclusion}, the paper is concluded with some remarks and suggestions for future work.

\section{The fundamental element: A single snapping shell}
\label{sec:SingleElement}

As a first step to understand the properties of the collective dimpled sheet, we consider the fundamental element \textit{i.e.} a single spherical cap embedded within a planar thin plate -- a `skirt'. (Alternatively, the sheet consists of `inclusions', the shells, embedded in a `matrix'. For a single element, we prefer to retain the term skirt as a reminder that it has a circular boundary, and to differentiate it from the matrix between multiple elements in a sheet.) We call this fundamental element a `snappit' (analogous to the bits used in electronic devices) and assume that its material and geometrical properties (save for the natural curvature) are identical to those of the remainder of the collective sheet; moreover, we shall assume that all the snappits of a dimpled sheet have the same characteristics. It should be noted that in this section we shall focus on the axisymmetric behaviours of a single element in \S2.1-\S2.4; we will turn to consider the causes of asymmetry in \S2.5.

\begin{figure}[ht]
\centering
\vspace{0.5cm}
\includegraphics[width=1.0\linewidth]{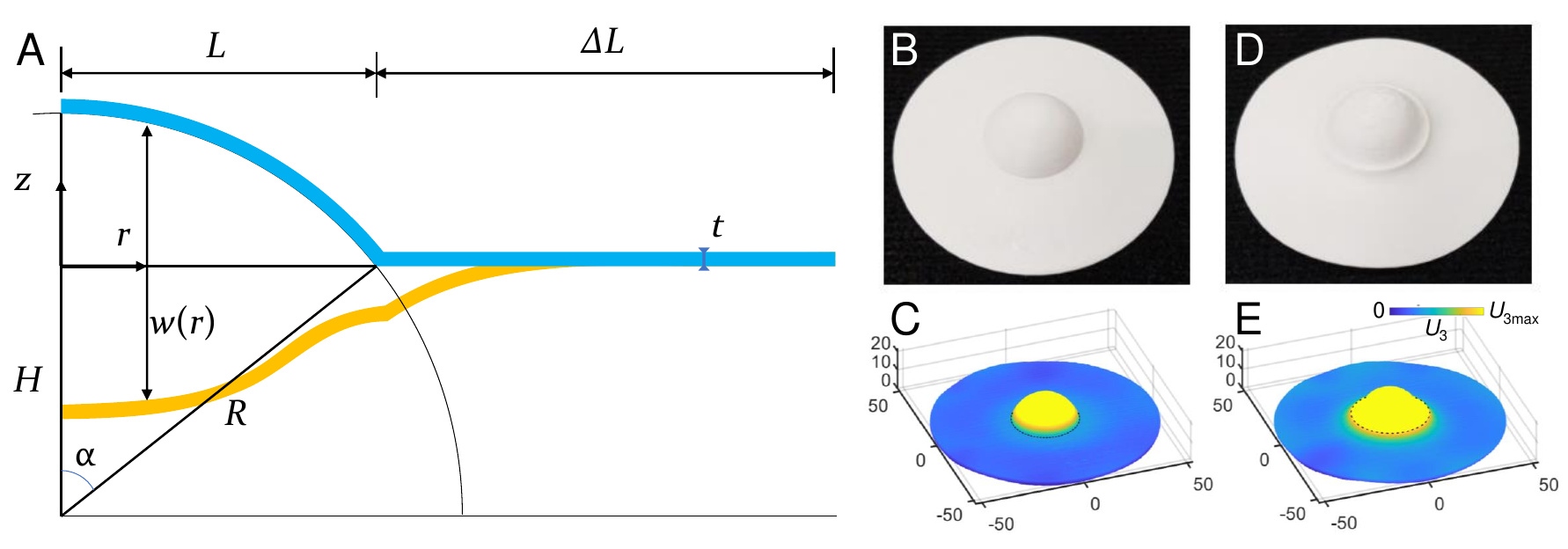}
    \caption{An individual snappit: (A) Geometrical parameters of a spherical cap with radius of curvature $R$, half-width $L$, thickness $t$, attached to a flat plate of thickness $t$, inner radius $L$, and outer radius $L+\dL$ (blue shape). When pushed in, the spherical cap is able to invert, deforming the outer plate in the process (yellow shape). (B)-(E) Photos and Scanned 3D profiles of the snappit in the natural configuration (B and C) and the inverted configuration (D and E). Here $L = 15\mathrm{~mm}$, $R = 18.75\mathrm{~mm}$, $t = 1\mathrm{~mm}$ and $\Delta L = 2L$ (corresponding to $\alpha = 0.93$, $\lambdad = 6.95$). Here $U_{3,\mathrm{max}} = 9.0$ and $8.0\mathrm{~mm}$ for C and E, respectively. Details of the experimental protocols are given in \ref{methods_exp}.}
\label{fig:ElementShapeFigure}
\end{figure}

The simplest geometry for a single snappit is to have an axisymmetric flat plate beyond the spherical cap, as shown in Fig.~\ref{fig:ElementShapeFigure}A. The geometrical parameters of this system are then given by those of the spherical cap (radius of curvature $R$, half-width $L$, thickness $t$) and of the flat skirt region (outer radius $L + \dL$ and, again, thickness $t$).

A typical example of this is shown in Fig.~\ref{fig:ElementShapeFigure}B and D: a 3D printed snappit can be popped from its natural state (see Fig.~\ref{fig:ElementShapeFigure}B and C) \cite[][]{Taffetani2018} to an inverted state (see Fig.~\ref{fig:ElementShapeFigure}D and E). This inversion induces a deformation of the plate that tends to be localized in the region near the interface with the shell.

Two natural questions emerge from this first picture of an inverted snappit: firstly, when is the inverted state stable (i.e.~when is the snappit bistable?) and secondly, what is the induced deformation in the sheet?

We consider each of these questions in turn and will use a combination of experiments, simulations and theory to tackle them: physical experiments using 3D printed samples, FEM simulations using commercial software (ABAQUS) and detailed theoretical analysis using shallow shell theory. With this we obtain a quantitative comparison between simulations and experiments, as well as analytical and asymptotic results.

\subsection{Stability of a shell: A review}
\label{sec:ShellStabilityReview}

To get an indication of the sort of behaviour that might be expected for a single snappit, we briefly review the scenario of a single spherical cap (without any extra material around the edge), which has been considered previously \cite[][]{Taffetani2018}. In this case the stability of the inverted state of a spherical cap is independent of the Young's modulus $E$ of the material: since there is no other force scale in the problem (assuming that gravity can be neglected) the transition between bistability and monostability is determined purely by the Poisson's ratio, $\nu$, and the geometrical properties of the shell, namely its radius of curvature $R$, half-width $L$, and thickness $t$.

On dimensional grounds, it is clear that with three length scales in the problem, two dimensionless groups can be formed. One natural choice is the angle sub-tended between the pole and the edge of the shell, i.e.~ $\alpha~ (=\sin^{-1}(L/R))$. Shells with $\alpha \ll 1$ are shallow \cite[][]{ventsel2001}, but with $\alpha = O(1)$ are deep. To determine a second dimensionless parameter, \cite{Taffetani2018} used energy arguments to consider the competition between bending and stretching energies: the bending energy density induced by inversion is $\mathcal{E}_B \sim B(1/R^2)$ (where $B=Et^3/12(1-\nu^2)$ is the bending stiffness of the shell) while the stretching energy density may be estimated to be $\mathcal{E}_S \sim Et\alpha^4$. The relative importance of stretching to bending energy densities is, therefore, $\mathcal{E}_S/\mathcal{E}_B=12(1-\nu^2)(R^2/t^2)\alpha^4$. We codify the relative importance of bending and stretching energies via the fourth root of this parameter \cite[][]{libai2005}, namely
\begin{equation}
\lambdad = [12(1-\nu^2)]^{1/4} \sqrt{\frac{R}{t}}\alpha.
\label{define_lambdad}
\end{equation}
Clearly the parameter $\lambdad$ involves both the depth of the shell, measured via $\alpha$, and its slenderness, $t/R$. However, \cite{Taffetani2018} found that the transition between monostability and bistability occurs at a critical value, $\lambdad=\lambda_\crit$ that is approximately independent of $\alpha$ for $\alpha\lesssim 1.0$ (from their results, the variation in $\lambda_\crit$ is less than $1\%$ for $\alpha\lesssim1.3$). Finally, \cite{Taffetani2018} showed that the value of $\lambda_\crit$ can be accurately calculated using shallow shell theory, which is only formally valid in the limit $\alpha\ll1$.

For the snappit with an attached outer skirt (see Fig. \ref{fig:ElementShapeFigure}), there is an additional parameter, namely the size of the skirt $\dL$ measured relative to the size of the shell itself, $L$; we therefore introduce the ratio
\begin{equation}
    \Delta=\dL/L.
    \label{eqn:DeltaDefn}
\end{equation} We now turn to characterize how the presence of a skirt affects the bistability of the snappit, as well as the overall shape in the inverted state. Ultimately, we will be interested in this variation as all three of the dimensionless parameters ($\lambdad$, $\alpha$ and $\Delta$) vary, but we initially simplify the problem by considering the problem for shallow shells, so that $\alpha\ll1$.

\subsection{Shallow shell formulation}
\label{sec:ShallowShellFormulation}

A first description of the shape of an inverted element was presented by \cite{sobota2020}; this involved employing a Rayleigh--Ritz approach with up to four degrees of freedom to solve the geometrically nonlinear shell model numerically. Here we provide some analytical insight by employing instead ``shallow-shell'' theory \cite[][]{calladine1989, ventsel2001}.

\paragraph*{Governing equations}\ 

Shallow-shell theory is based on the equations of axisymmetric plate theory modified to incorporate the finite radius of curvature of the shell. As a result, the same equations can be used to describe the vertical (normal) deflection, $w(r)$, and stress potential, $\psi(r)$, in a cylindrical polar geometry for both the shell and the skirt region if we introduce a spatially varying natural curvature $$\kappa(r)=R^{-1} H(L-r),$$ where $H(\cdot)$ is the Heaviside step function. We therefore have
\begin{equation}
B\nabla^4 w + \kappa(r)\frac{1}{r}\frac{\upd}{\upd r}(r \psi)-\frac{1}{r}\frac{\upd}{\upd r}\left(\psi \frac{\upd w}{\upd r}\right)=0,
\label{Shell_equation_1}
\end{equation}
and
\begin{equation}
\frac{1}{Y}r\frac{\upd}{\upd r}\left[\frac{1}{r}\frac{\upd}{\upd r}(r \psi)\right] = \kappa(r)r\frac{\upd w}{\upd r} - \frac{1}{2}
\left(\frac{\upd w}{\upd r} \right)^2,
\label{Shell_equation_2}
\end{equation} where $Y=Et$ is the stretching modulus of the material and the stress potential $\psi$ is the derivative of the Airy stress function (defined such that $\sigma_{rr}=\psi/r$ and $\sigma_{\theta \theta}=\upd \psi/\upd r$).

\paragraph*{Non-dimensionalization}\ 

We follow \citet{Taffetani2018} by introducing dimensionless variables
\begin{align*}
\rho = \frac{r}{L}, \quad W(\rho) = w(r)\frac{R}{L^2} \quad \mathrm{and} \quad \Psi(\rho) = \psi(r)\frac{R^2}{YL^3},
\end{align*} so that the governing equations \eqref{Shell_equation_1}-\eqref{Shell_equation_2} become
\begin{equation}
\lambda_s^{-4} \nabla^4 W + \frac{\mathcal{S}(\rho)}{\rho}\frac{\upd}{\upd\rho}(\rho \Psi)-\frac{1}{\rho}\frac{\upd}{\upd\rho}\left(\Psi \frac{\upd W}{\upd\rho}\right)=0,
\label{Scaled_Shell_equation_1}
\end{equation}
and
\begin{equation}
\rho \frac{\upd}{\upd\rho} \left[\frac{1}{\rho}\frac{\upd}{\upd\rho}(\rho \Psi)\right] = \mathcal{S(\rho)}\rho\frac{\upd W}{\upd\rho} - \frac{1}{2}
\left(\frac{\upd W}{\upd\rho} \right)^2
\label{Scaled_Shell_equation_2}
\end{equation} where $\mathcal{S}(\rho)=H(1-\rho)$ is the indicator function for the shell region (i.e.~$\mathcal{S}=1$ in the shell region, $0 < \rho < 1$, and $\mathcal{S}=0$ in the skirt region, $1<\rho < 1+\Delta$). This non-dimensionalization introduces two dimensionless parameters: $\Delta$ as expected, and 
\begin{equation}
\lambda_s = \left[12 \left(1-\nu^2\right)\right]^\frac{1}{4}\frac{L}{(tR)^{1/2}}.
\label{Parameter_lambda}
\end{equation} (Note that $\lambdas$, which characterizes the geometry of the shell, is the shallow shell version of the parameter, i.e.~$\lambdas\approx\lambdad$ for $\alpha\ll1$.) 

To be able to solve equations \eqref{Scaled_Shell_equation_1}--\eqref{Scaled_Shell_equation_2}, we must specify appropriate boundary conditions at the boundaries, $\rho=0$ (shell centre), $1+\Delta$ (skirt edge) and, crucially, the join between the shell and skirt regions, $\rho=1$.

\paragraph*{Boundary conditions}\

The appropriate boundary conditions at the shell centre, $\rho=0$, and skirt outer edge, $\rho=1+\Delta$, are relatively straightforward: at the shell centre, we use the symmetry and no displacement conditions $W'(0)=W(0)=U_r(0)=0$, where $U_r$ is the horizontal (radial) displacement. (Note that since the governing equations only involve $W'$, not $W$ itself, there is some freedom in the choice of $W(0)$.) At the outer boundary of the skirt, $\rho = 1 + \Delta$, we assume there is zero bending moment, zero shear force and no radial stress ($\sigma_{rr}=0$).

The interface between the shell and skirt region at $\rho = 1$ deserves further discussion. We have immediately that the horizontal and vertical displacements are continuous here, as is the radial stress, so that $[W]_{1^-}^{1^+}=[U_r]_{1^-}^{1^+}=[\sigma_{rr}]_{1^-}^{1^+}=0$. It is also natural to assume that the bending moment and out-of-plane shear force are continuous. (The in-plane shear force, $\sigma_{r\theta}=0$ from our assumption of axisymmetry.) The slope of the deflection at the boundary may, in principle, be discontinuous (depending on how this join is implemented); however, we assume that this join is rigid so that the angle between the tangent to the shell at its edge and the tangent to the skirt remains constant at $\alpha$. Given the difference in natural shapes, this means that the slope of the displacement is continuous, i.e.~$[W']_{1^-}^{1^+}=0$. The boundary conditions discussed above are summarized in terms of $W$, $\psi$ and their derivatives in Table \ref{tab:Boundarycondition} of  \ref{sec:AppB3}.

The Eqs. \eqref{Scaled_Shell_equation_1}--\eqref{Scaled_Shell_equation_2} can be solved subject to these boundary conditions using, for example, the multipoint boundary value problem solver \texttt{bvp4c} in MATLAB. We shall discuss the behaviour of the inverted solutions of these equations in Section \ref{sec:DeformedElementShape} but first focus on determining when this inverted shape exists numerically, i.e.~the critical condition for the bistability of the snappit. Note that this numerical system is controlled by just two dimensionless parameters, $\Delta$ and $\lambdas$, since the shallow shell formulation assumes that the shells are shallow, \textit{i.e.}~$\alpha\ll1$.

\subsection{Bistability}
\label{sec:Bistability}

\begin{figure}[ht]
    \centering
    \vspace{0.5cm}
    \includegraphics[width=0.85\linewidth]{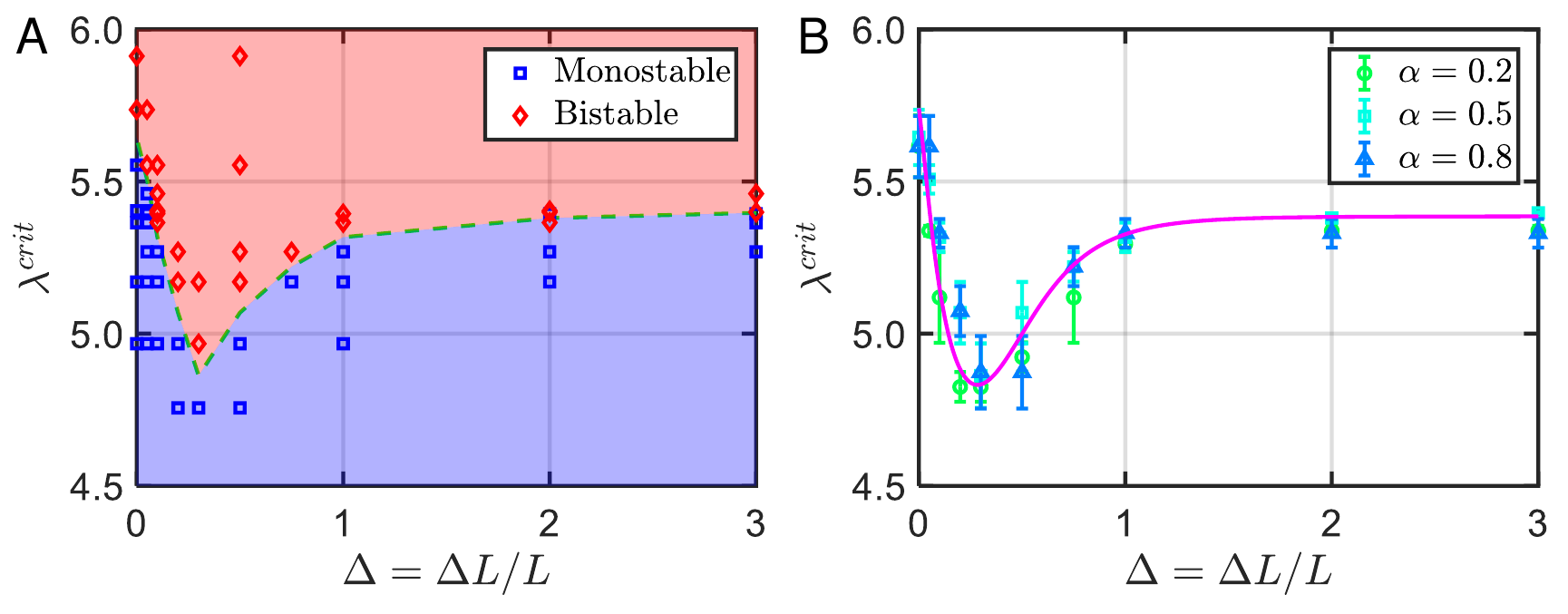}
    \caption{Bistability of the single snappit (or shell-skirt element). (A) Phase diagram characterizing the stability of an element with $\alpha = 0.52$ as the shell depth parameter, $\lambdad$, and the relative plate size, $\Delta = \dL/L$, vary. The symbols are obtained from FEM simulations, in which red diamonds represent the bistable cases, and blue squares represent the monostable cases. The phase boundary, calculated as the mean of the limiting cases, is indicated by the green dashed curve. (B) The transition from monostable to bistable as determined from FEM simulations with different values of $\alpha$ (points) agrees well with the theoretical prediction of shallow shell theory (pink solid curve). Circular, square and triangular symbols correspond to the transition from bistable to monostable for cases with $\alpha$ = 0.20, 0.52 and 0.80, respectively.}
    \label{fig:Bistability}
\end{figure}

Our first goal is to understand the geometrical conditions under which a single snappit is bistable, i.e. when does the inverted shape actually exist without the application of any external loads? As already discussed, there remain only three dimensionless parameters in the problem ($\alpha$, $\lambda$ and $\Delta$), together with the material property $\nu$. We present two approaches to determine the critical condition for bistability here: firstly, we use FEM simulations (see \ref{methods_fem} for details) with a range of values of $\alpha$, $\lambdad$ and $\Delta$ and determine whether the inverted state is stable (and hence the snappit is bistable). This gives a rough indication of the behaviour of the critical value $\lambdad^\crit(\Delta,\alpha;\nu)$; our results are shown in Fig.~\ref{fig:Bistability}A, with the dependence of the threshold $\lambdad^\crit$ as a function of $\Delta$ shown in Fig.~\ref{fig:Bistability}B. Note that here we determine the threshold, $\lambdad^\crit$, as the mean of the smallest $\lambdad$ with bistable behaviour and the largest $\lambdad$ with monostable behaviour for given $\Delta$; error bars are then used to record the difference between these two values. Secondly, we use arc-length continuation in AUTO \cite[][]{doedel2007} to solve the shallow shell system \eqref{Scaled_Shell_equation_1}--\eqref{Scaled_Shell_equation_2} and determine the critical value $\lambdas^\crit(\Delta)$ at which the inverted state ceases to exist. Our continuation calculation involves three key steps: first, the perfectly everted solution is used as an initial guess to the solution of the system \eqref{Scaled_Shell_equation_1}--\eqref{Scaled_Shell_equation_2}  everted solution for an arbitrary, but large, value of $\lambdas$ and $\Delta L/L$ = 3. The true solution is `close' to the everted solution for large $\lambdas$, and so the true solution can readily be found using a relaxation method. Secondly, we perform a one parameter continuation analysis, reducing the value of $\lambda_s$, but holding  $\Delta L/L=3$ constant. This allows the branch of stable everted solutions to be followed until a fold bifurcation in the system is reached. The position of the fold bifurcation in terms of $\lambdas$ corresponds to the $\lambda^{\crit}$ at the monostable-bistable threshold for the given $\Delta L/L$ = 3. Finally, we perform a two-parameter continuation, following the fold while varying $\Delta L/L$ and $\lambdas$ accordingly. This allows us to extract the value of $\lambda^{\crit}$ for each $\Delta L/L$ down to $\Delta L/L=0$. This calculation gives the solid curve in Fig.~\ref{fig:Bistability}B, which agrees very well with the FEM results.

On the whole, Fig.~\ref{fig:Bistability} shows that the presence of a skirt decreases the threshold value of $\lambdad$ at which the shell becomes bistable: compared with the case of a spherical cap without the skirt ($\lambda^{\crit}\left(\Delta=0\right) \approx 5.75$), the presence of a skirt region makes the element `more' bistable because, for a given $\Delta$, all the geometrical configurations with $\lambdad$ in between $\lambda^{\crit}(\Delta)$ and $\lambdad \approx 5.75$ are now within the bistable region. An interesting feature of the behaviour seen in the phase diagram (Fig.~\ref{fig:Bistability}A) is that, while there is some dependence of the critical shell depth $\lambdad$ for bistability on the size of the plate region, $\Delta$, this dependence is (perhaps surprisingly) small: over a wide range of $\Delta$, the critical value of $\lambdad$ changes by less than $10\%$. We also see that, as observed previously \cite[][]{Taffetani2018}, the effect of the angle $\alpha=\sin^{-1}(L/R)$ on the bistability is quite limited (indeed, it is not detectable in the results presented in Fig.~\ref{fig:Bistability}B). As a result, we shall follow \citet{Taffetani2018}: we use shallow shell theory to understand the importance of the parameter $\lambda=\lambdas$, and apply these results to less shallow shells (with $\alpha\lesssim1$, not $\alpha\ll1$) simply by letting $\lambda=\lambdad$.

\subsection{Shape of an axisymmetric deformed element}
\label{sec:DeformedElementShape}

\begin{figure}[ht]
    \centering
    \vspace{0.5cm}
    \includegraphics[width=.80\linewidth]{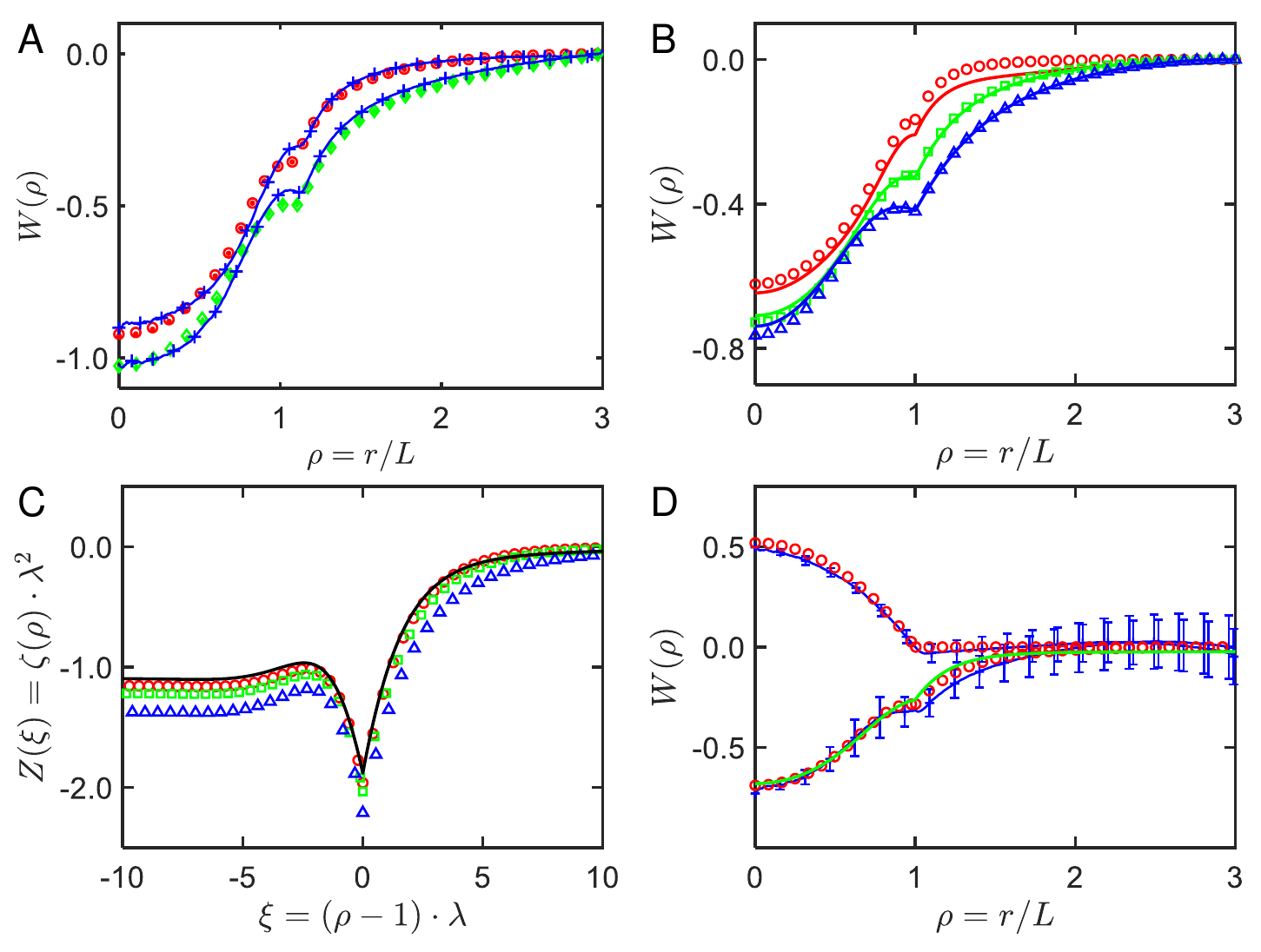}
    \caption{The deformed shape of a single snappit. (A) Comparison between experimental measurements of the deformed, inverted shape (blue solid lines with plus sign) and two sets of FEM simulations with linear elastic (larger open symbols) and hyperelastic (smaller filled symbols) material properties. Results are shown for two values of $\lambdad$: $\lambdad = 8.63$ (red circles) and $\lambdad = 7.05$ (green diamonds), but constant $\alpha = 1.08$ and $\Delta = 2$. (B) Comparison between FEM simulations (open symbols) and the predictions of shallow shell theory (solid curves) with $\lambda=\lambdad=13.32$ (red), $7.69$ (green) and $5.96$ (blue), respectively. (C) The boundary layer scaling collapses FEM simulations with different values of $\lambdad\gg1$ (open symbols, with $\lambdad=13.32$, circles, $24.31$, squares, and $42.11$, triangles) as $\lambdad\to\infty$; moreover, these FEM simulations tend to the solution of the boundary layer problem based on the shallow shell assumption, found by solving \eqref{BL_shelleq1}--\eqref{BL_shelleq2} (solid curve). (D) Comparison between experiment (solid blue line with plus sign and error bar), FEM simulation (open symbols) and the boundary layer prediction (solid green line) for a shallow shell with $\lambdad=7.69$. In (B)-(D), we fix $\alpha = 0.38$ and $\Delta = 2$.}
    \label{fig:ElementShape}
\end{figure}

Having determined when the element is bistable, we now move on to understand the shape in the inverted state. Given the success of shallow shell theory in describing the transition from mono- to bi-stable just demonstrated, it would be convenient to be able to use shallow shell theory to study in detail the deformed shape. First, however, we should confirm whether the effect of nonlinear elasticity is significant experimentally. In Fig.~\ref{fig:ElementShape}A we compare results from FEM simulations (see \ref{methods_fem} for details), both using hyperelastic and linearly elastic constitutive models with results from physical experiments (see \ref{methods_exp}). We see first that FEM results with each constitutive law are indistinguishable at the scale of the plot and, further, that either constitutive law does an equally satisfactory job of describing the experimental data. We shall therefore continue to study the shape assuming a linearly elastic model and, since we have also seen that the parameter $\alpha$ has a limited effect on the transition from mono- to bistability if we identify $\lambdad\rightarrow\lambdas$, we use shallow shell theory. The numerical solution of the shallow shell equations gives a good account of the simulation results (see Fig.~\ref{fig:ElementShape}B); crucially, however, we will be able to gain some analytical insight in the limit of thin shells $\lambdas\gg1$, even though shallow shell theory is formally only valid for $\alpha\ll1$.

\paragraph*{Boundary layer analysis for $\lambdas\gg1$}\

Both our FEM simulations and numerical solutions of the shallow shell equations suggest that the deformation from the perfectly inverted shape becomes increasingly localized to the boundary between skirt and shell, at $\rho=1$, as $\lambdas$ increases (see Fig.~\ref{fig:ElementShape}B). We see from \eqref{Scaled_Shell_equation_1} that as $\lambdas \to \infty$ the bi-Laplacian term becomes less important, except possibly in small regions where the derivatives become large --- this term may therefore be neglected except in spatial boundary layers, which we expect to be centred on $\rho=1$. Moreover, the `outer' solution in the majority of the shell corresponds simply to ``perfect eversion'' within the shell region, $W\sim\rho^2-1$, and a constant vertical displacement within the skirt region, $W\sim\mathrm{cst}$; this deformation also results in zero stress within the majority of the shell. In this limit, therefore, the deformation is localized within the boundary layer, and the size of the boundary layer gives the scale over which the snappit deformation is expected to decay. We shall see that this length scale is $\sim \lambda^{-1}$, which in physical variables corresponds to the Pogorelov length scale, $\lp\sim(tR)^{1/2}$, as seen in many other problems involving the inversion of spherical caps \cite[][]{libai2005,Pogorelov1988,Gomez2016}. 

Firstly, we can decompose the displacement of the snappit as a perfect inverted shape plus a small perturbation, i.e.
\begin{equation} 
W(\rho) = \mathcal{S}(\rho)(\rho^2-1) + \zeta(\rho)
\label{Shell_perturbation}
\end{equation}

For the moment, however, we proceed generally by letting
\begin{align}\begin{aligned}
\rho &= 1+\lambdas^{-\mu} \xi\\
\zeta(\rho) &= \lambdas^{-\beta} Z(\xi)\\
\Psi(\rho) &= \lambdas^{-\gamma} \varPsi(\xi).
\label{BL_variable}
\end{aligned}\end{align} 
Examining the transformed \eqref{Scaled_Shell_equation_1} in the plate region, and requiring terms to balance, we find that:
\begin{equation}
4(1-\mu) = \gamma - 2 \mu
\label{balanceeq1}
\end{equation}
while from \eqref{Scaled_Shell_equation_2} we have:
\begin{equation}
\gamma = 2\beta.
\label{balanceeq2}
\end{equation}
Combining this with \eqref{balanceeq1}, we therefore find that $\mu + \beta = 2$. To make any further progress requires information from the shell region; since this information only enters through the $\rho \zeta'$ term in \eqref{Scaled_Shell_equation_2}, we require this to enter at the same order as the other terms so that $\mu=\beta$, and hence $\mu=\beta = 1$, $\gamma = 2$.

Plotting the results of FEM simulations in terms of the boundary layer variables suggested by this analysis (specifically, plotting $Z(\xi) = \zeta(\rho)\times \lambdas$ as a function of $\xi = \lambdas(\rho - 1) $), shows a good collapse as $\lambdas \rightarrow \infty$ (see Fig.~\ref{fig:ElementShape}C). To understand the limiting behaviour, i.e.~with $\lambdas=\infty$, we note that \eqref{BL_variable} transforms \eqref{Scaled_Shell_equation_1} and \eqref{Scaled_Shell_equation_2} to
\begin{equation}
\frac{\upd^4 Z}{\upd \xi^4} - \mathcal{S}(\rho) \frac{\upd \varPsi}{\upd \xi} - \frac{\upd}{\upd \xi} \left(\varPsi \frac{\upd Z}{\upd \xi}\right) = 0
\label{BL_shelleq1}
\end{equation}
and
\begin{equation}
\frac{\upd^2 \varPsi}{\upd \xi^2} = -\mathcal{S}(\rho) \frac{\upd Z}{\upd \xi} -\frac{1}{2} \left(\frac{\upd Z}{\upd \xi} \right)^2.
\label{BL_shelleq2}
\end{equation} These equations are to be solved subject to the condition that displacement and stress should vanish away from the join region; within the shell region we take $Z_s',Z_s'',\varPsi_s\to0$, as $\xi \rightarrow -\infty$ (since the natural shell curvature has already been subtracted). Within the skirt region we take $Z_p,Z_p',\varPsi_p\to0$, as $\xi \rightarrow \infty$, setting zero vertical displacement at infinity to fix the object in space. At the interface between shell and skirt, we also require continuity of these fields as well as the horizontal displacement i.e. $0=Z_s-Z_p=Z_s'-Z_p'=Z_s''-Z_p''=Z_s'''-Z_p'''$ and $0=\varPsi_s-\varPsi_p=\varPsi_s'-\varPsi_p'$ at $\xi = 0$. Solving \eqref{BL_shelleq1}--\eqref{BL_shelleq2} numerically subject to these condition yields the solid black curve shown in Fig.~\ref{fig:ElementShape}C. Note that this boundary layer solution is in good agreement not only with the results of FEM simulations (Fig.~\ref{fig:ElementShape}C), but also agrees well with experiments when cast back into the shape variables (Fig.~\ref{fig:ElementShape}D).

This analysis shows that the join between the skirt and shell is the cause of the deformation observed in an inverted snappit: there is an incompatibility between the perfectly inverted shape (an inverted spherical cap joined to a flat skirt region) and the (fixed) angle between the shell and the flat region at the join. This geometrical origin of the boundary layer is distinct to the boundary layer region in an inverted spherical cap \cite[][]{libai2005,Taffetani2018}, which is caused by a mismatch between the zero applied moment condition and the finite, but constant, bending moment required to maintain the inverted spherical shape.

As well as showing that the join region is what is responsible for the deformation of the skirt region, the boundary layer analysis also gives us insight into the structure of the stress profile within the deformed snappit. In particular, the boundary layer scaling gives $\sigma_{\theta \theta}(=\upd\Psi/\upd\rho) = O(\lambdas^{-1})$ while $\sigma_{rr}(=\Psi/\rho) = O(\lambdas^{-2})$; as a result, we expect that $\sigma_{\theta\theta}\gg\sigma_{rr}$, which is consistent with the numerical results. We also observe that both of these stresses are negative at some point within the vicinity of the join between the sheet and shell regions: the snappit is under both radial and azimuthal compression. In general, such a compression may be expected to lead to a buckling instability, and so we turn now to consider this instability. 

\subsection{Origin of instability}
\label{sec:Origin_instability}

\begin{figure}[ht]%[tbhp]
    \centering
    \vspace{0.5cm}
    \includegraphics[width=0.7\linewidth]{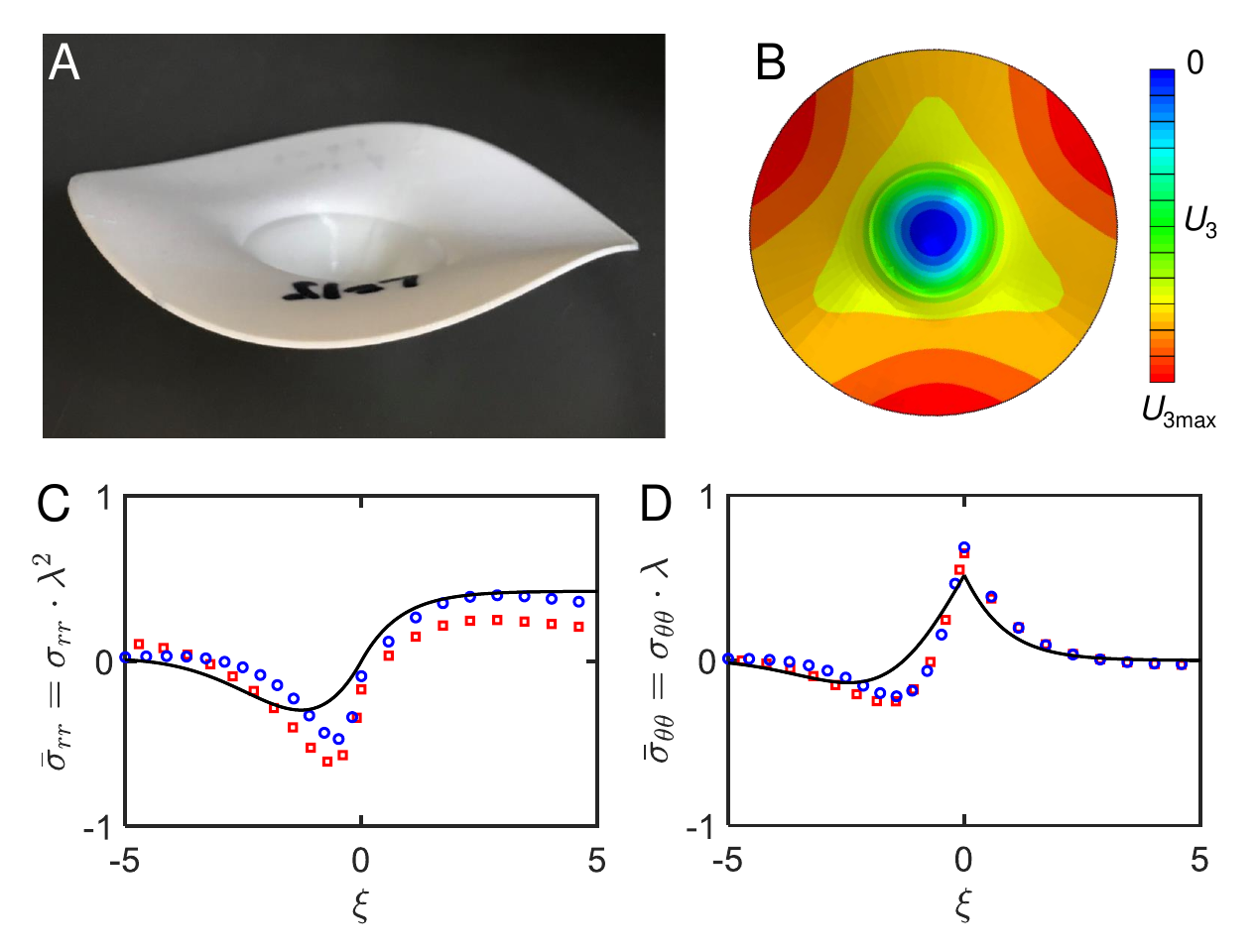}
    \caption{Azimuthal buckling of a single snappit. (A) Experimental observation and (B) FEM prediction for the vertical displacement, $U_{3}$, of the buckling of an element with $\alpha=1.08$, $\Delta=2$ and $\lambdad = 8.63$; here $U_{\rm 3max} = 27.1~{\rm mm}$. The distribution of the scaled (C) hoop stresses ($\bar{\sigma}_{\theta \theta} = \sigma_{\theta \theta} \cdot \lambda$, blue symbols) and (D) radial stresses ($\bar{\sigma}_{rr} = \sigma_{rr} \cdot \lambda^2$, red symbols) stresses within the snapped elements with $\lambdad = 8.63$ (squares) and $\lambdad = 17.26$ (circles) as calculated from FEM simulations, plotted as a function of $\xi=(\rho-1)\lambda$. The corresponding predictions of the boundary layer theory (solid curves in both (C) and (D)) show that the compressive hoop stress $|\sigma_{\theta\theta}|=O(\lambda^{-1})$, and hence is significantly larger than the typical compressive radial stress $|\sigma_{rr}|=O(\lambda^{-2})$; this explains the azimuthal, rather than radial, buckling instability observed here.
    }
    \label{fig:ElementBuckling}
\end{figure}

The analyses of Sections \ref{sec:Bistability} and \ref{sec:DeformedElementShape} were predicated on the assumption that the deformation remains axisymmetric. However, in some scenarios, a visible azimuthal instability occurs (see Fig.~\ref{fig:ElementBuckling}A,B). It is natural to treat this as a buckling instability and the boundary layer theory just presented gives two insights: firstly, both the hoop and radial stresses are negative (compressive) in some region close to the join between sheet and shell, see Fig.~\ref{fig:ElementBuckling}C,D. Secondly, in the axisymmetric state, the hoop stress is an order $\lambdas$ larger than the radial stress (from Fig.~\ref{fig:ElementBuckling}C, $\sigma_{rr} \sim \lambda_s^{-2}$, while from Fig.~\ref{fig:ElementBuckling}D, $\sigma_{\theta\theta}\sim \lambda_s^{-1}$); therefore we might expect relieving the hoop compression by azimuthal buckling to be more energetically favourable than relieving the radial compression by buckling in the radial direction.

While this qualitative discussion explains the origin of the instability, to go further would require a more detailed analysis, along the lines of standard post-buckling analyses. On the face of it, the fundamental problem we consider has some similarities with the Lam\'{e} problem \cite[][]{Coman2007,Davidovitch2011} in which an annular sheet is subject to different internal and external tensions; in particular, here an annular sheet (the skirt) is subject to a radial displacement at its inner edge and suffers an azimuthal instability as a result.

However, there are some important differences with the standard Lam\'{e} problem. Most notably, in the Lam\'{e} problem, there are two dimensionless parameters: one corresponding to the ratio of the applied tensions (or imposed displacements) and the other related to the ease with which the sheet bends (the `bendability' defined by \cite{Davidovitch2011}). As a result, for a fixed bendability one can consider gradually increasing the tension ratio until a small, buckled perturbation with a preferred wavenumber appears (as an eigenfunction of the problem). This is a `Near threshold' analysis. In contrast, in the inversion-induced instability we report here, there is only a single dimensionless parameter ($\lambda$), which encodes both how easily bent the sheet is \emph{and} the magnitude of the inward displacement that occurs at the inner edge. In some sense, the difference arises because there is no external tension applied at the outer boundary. This is similar to the case of an elastic ring subject to an internal tension, but no external tension, which has been shown to buckle with a wavelength $\lambda\sim(B/TR_{\mathrm{in}}^2)^{1/2}$ \cite[][]{Box2020}, which would correspond to a mode number of instability $n\sim\lambda$. However, the instability studied by \citet{Box2020} is driven by the dynamics of buckling and cannot occur statically --- this is therefore not relevant to the static problem considered here. Instead, it seems that the mode number of instability is determined in some other way, as discussed in the tensionless Lam\'{e} problem by \cite{Pal2022}. We do not consider this problem further here, leaving it for the future, but focus instead on how the system collectively chooses to relieve compression.

\section{Many elements: Global response to distributed compression}
\label{sec:ManyElements}

The response of a single skirt region to the azimuthal compression caused by the presence of an inverted dimple is naturally to buckle azimuthally. When multiple elements are combined in a single sheet, i.e.~the skirt region is shared between multiple elements to become a matrix, it is not  clear what the global response to compression should be. We therefore begin by studying experimentally and numerically the response of the smallest symmetric system that contains multiple elements embedded within a circular sheet: a single snappit surrounded by six (hexagonally packed) other snappits. In this case, the snappits form a triangular lattice.

\subsection{A triangular lattice}

\begin{figure}[ht]%[tbhp]
    \centering
    \vspace{0.5cm}
    \includegraphics[width=1.0\linewidth]{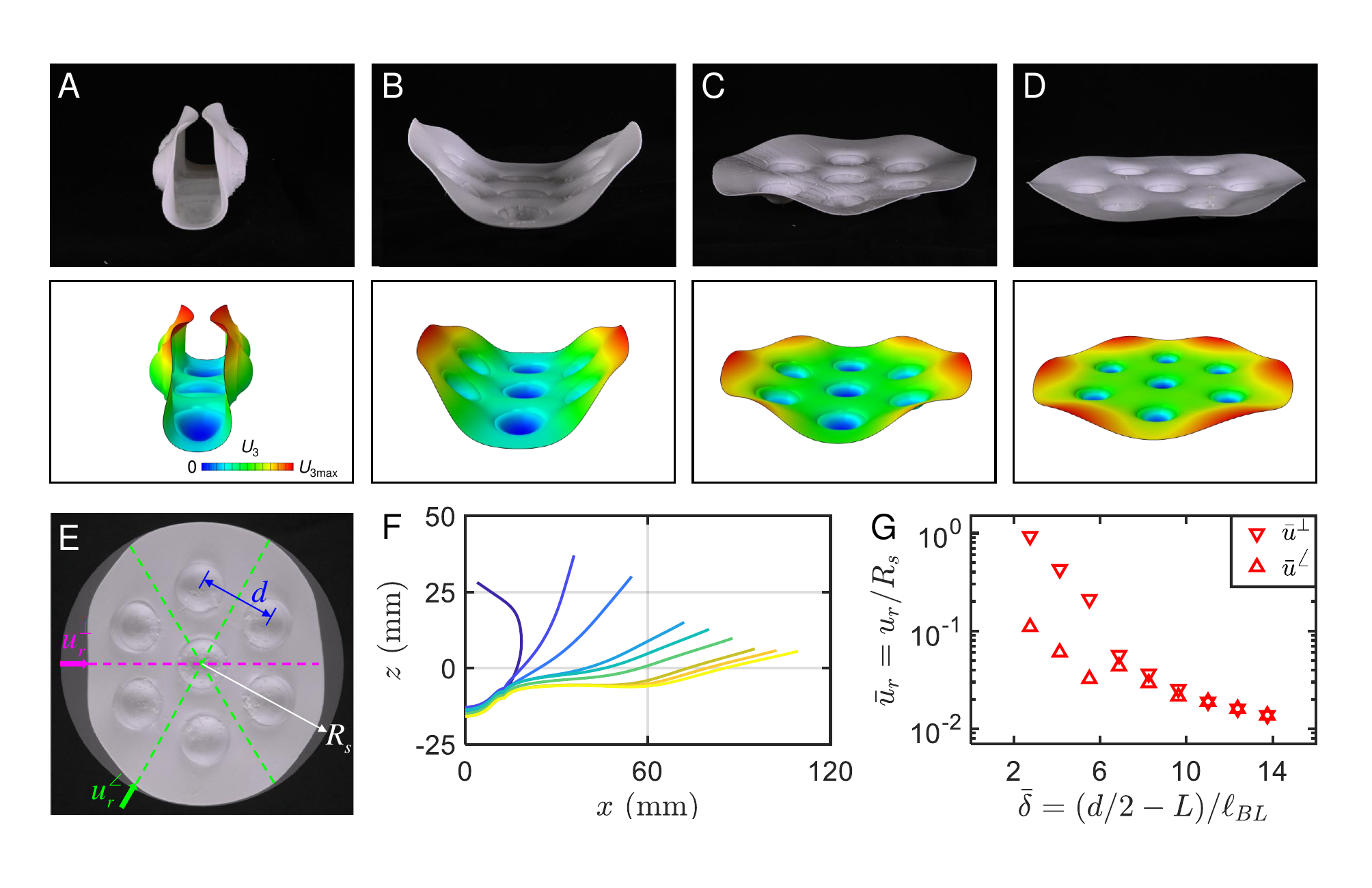}
    \caption{Demonstration of the shape-morphing of a soft dimpled sheet with snappits arranged on a triangular lattice. (A-D) Experiments (top row) and FEM simulations (bottom row) of snapped dimpled sheets (fixing $\alpha = 1.08$, $\lambdad = 8.25$, $L=12\mathrm{~mm}$) with different dimple spacing, from (A) to (D), $d = $32, 40, 44 and 56$\mathrm{~mm}$, in which, $U_{\rm 3max} = 53.97,~51.80,~30.09,~29.35~{\rm mm}$, respectively. (E) Overlaid plan views of the original (faint) and snapped (foreground) shape of a dimpled sheet with radius $R_s = \sqrt{3}d$ and $d = 40$ mm. Here, the snapped state is curved along a major axis. The magenta and green dashed lines show the direction in which displacements perpendicular ($\perp$) and inclined ($\angle$)  to the major curved axis are measured. (F) The profiles of snapped sheets along the perpendicular axis (i.e.~the magenta line in panel E) as the dimple spacing, $d$, varies in the range $32\mathrm{~mm}\leq d\leq56\mathrm{~mm}$. (G) The normalized horizontal displacement of  snapped  sheets     $\bar{u}_r^\perp$ and $\bar{u}_r^\angle$ (as marked in panel E) as  functions of the normalized dimple spacing, $\bar{\delta}=(d/2-L)/\ell_{BL}$. Note that for the cases shown in (A)-(D), $\bar{\delta}=$ 2.75, 5.50, 6.88 and 11.00, respectively.}
    \label{fig:GlobalMorphingDemonstration}
\end{figure}

The fundamental geometry is shown in Fig.~\ref{fig:GlobalMorphingDemonstration}E. This geometry is characterized by the properties of each, identical, snappit (i.e.~its thickness, radius of curvature and base width) but also the separation between the centres of each snappit, $d$. We focus here on the effect of the snappit separation, $d$, on the behaviour of the whole sheet. Heuristically, we might expect that if the separation between snappits is `large' then they are effectively isolated with no significant interaction between them. Moreover, our analysis of the single snappit suggests that the boundary layer thickness $\ell_{BL}=L /\lambda$ is the natural horizontal length scale over which the deformation induced by each snappit decays. Fixing $\lambda$, we therefore focus on the effect of the dimensionless gap width between snappit edges:

\begin{equation}
    \bar{\delta}=(d/2-L)/\ell_{BL}.
    \label{eqn:DeltaDefn}
\end{equation} Defined in this way, $\bar{\delta}$ measures the half-spacing between two snappits in terms of the boundary layer width surrounding each snappit; we therefore expect that for $\bar{\delta}$ sufficiently small the interaction between snappits should be strong, while for sufficiently large $\bar{\delta}$ the snappits are effectively isolated.

Figs.~ \ref{fig:GlobalMorphingDemonstration}A-D show comparisons between the shapes observed experimentally and in FEM  simulations as the distance between snappits changes. The shapes predicted by FEM simulations agree well with those observed experimentally. As expected, both approaches show that as the separation between snappits increases the collective deformation of the sheet decreases. More interestingly, however, we see that it is not just the magnitude of deformation  observed that changes with $\delta$, but also the \emph{type} of deformation: for small $\bar{\delta}$ the sheet deforms cylindrically (forming a taco-like shape, as in Fig.~\ref{fig:GlobalMorphingDemonstration}A) while for large $\delta$ the shape is (approximately) axisymmetric, see Fig.~\ref{fig:GlobalMorphingDemonstration}D for example. More  specifically, in Fig. ~\ref{fig:GlobalMorphingDemonstration}F we plot profiles, taken along the dashed magenta line, of snapped dimpled sheets with  dimple spacing $d$ varying in the range  $32\mathrm{~mm}\leq d\leq56\mathrm{~mm}$ . It is clearly seen that decreasing $d$ is associated with a significant change in shape. We claim that the appearance of this buckling transition is crucial in the appearance of soft modes in the soft shapeable sheet: on a triangular lattice, the cylindrical mode has a multiplicity of three (the fold can happen along each of the three axes) and so multiple global shapes can be formed in a larger sheet. 

To understand this global transition in shape, it is natural to draw analogies to other systems in which out-of-plane deformation occurs as a result of in-plane strain. While this is the fundamental origin of Euler buckling, two more specific examples are worthy of some discussion: deformation caused by inhomogeneous growth or shrinkage in a thin, single-layer material \cite[][]{Klein2007,Dervaux2008,Sharon2010} and that caused by differential thermal expansion in a bilayer material \cite[][]{Freund2006}. In both cases, deforming out of plane relaxes some of the in-plane strain when the system is flat, but the means of doing so, and conditions under which it occurs, are quite different. When the strain profile through the sheet thickness is uniform a radial inhomogeneity in growth/shrinkage must be introduced to obtain buckled shapes; moreover, bowl-like shapes (positive Gaussian curvature) are obtained from increasing shrinkage at the edge while azimuthally oscillating shapes (negative Gaussian curvature) are observed with increased shrinkage at the centre \cite[][]{Klein2007}. In contrast, a bilayer with differential strains in the two layers adopts a bowl-like shape (positive Gaussian curvature) before transitioning to a more cylindrical shape (though still with positive Gaussian curvature) above a critical strain \cite[][]{Freund2006}.

It is natural to assume that reducing the spacing between identical snapping elements (i.e.~reducing $\bar{\delta}$) corresponds to increasing the level of strain in the material.  Given this, the qualitative behaviour observed as $\bar{\delta}$ varies in  a dimpled sheet is essentially the same as that observed in the bilayer sheet: Fig.~ \ref{fig:GlobalMorphingDemonstration} shows that sheets with largest $\bar{\delta}$ appear to be rotationally symmetric in the inverted state, while those with smallest $\bar{\delta}$ clearly break the underlying rotational symmetry. This similarity to a bilayer sheet might be rationalized by the observation that the contraction applied by the individual dimples lies below the neutral plane of the unstrained parts of the matrix region between them.

To quantify the transition from rotational symmetry to two-fold symmetry, we  measure the horizontal displacement perpendicular to the fold that ultimately forms (denoted $u_r^\perp$), as well as the horizontal displacement measured at $60^\circ$ to this axis (denoted $u_r^\angle$) --- see sketch in Fig.~\ref{fig:GlobalMorphingDemonstration}E for these definitions. (Quantifying the curvature in different directions is difficult because of the dimples' shape; we argue that measuring $u_r^\perp$ and $u_r^\angle$ is both more reproducible and clearer.) Plotting these displacements as a function of $\bar{\delta}$, see Fig.~\ref{fig:GlobalMorphingDemonstration}G, shows a sharp transition as $\bar{\delta}$ decreases: for $\bar{\delta}\gtrsim 10$, we have $u_r^\perp\approx u_r^\angle$ (see Fig. ~\ref{fig:GlobalMorphingDemonstration}D), while for $\bar{\delta}\lesssim 6$, $u_r^\perp\gg u_r^\angle$ (Figs. ~\ref{fig:GlobalMorphingDemonstration}A and B); there is also a notable intermediate range, i.e., $6 \lesssim \bar{\delta} \lesssim 10$, in which $u_r^\perp$ is slightly larger than $u_r^\angle$; this corresponds to the slightly curved state shown in Fig.~\ref{fig:GlobalMorphingDemonstration}C.  We also note that this transition around $\bar{\delta}\approx6$ corresponds to the deformation and stress profiles within the two boundary layers of the interacting snappits no longer overlapping:  from Fig.~\ref{fig:ElementShape}C and Figs. \ref{fig:ElementBuckling}C and D we see that the effects of the boundary layer are localized within a region $|r-L|\lesssim5\ell_{BL}$ $(|\xi|\lesssim5)$ of the join between element and skirt.

\subsection{A hexagonal lattice}

\begin{figure}[ht]
    \centering
    \vspace{0.5cm}
    \includegraphics[width=1.0\linewidth]{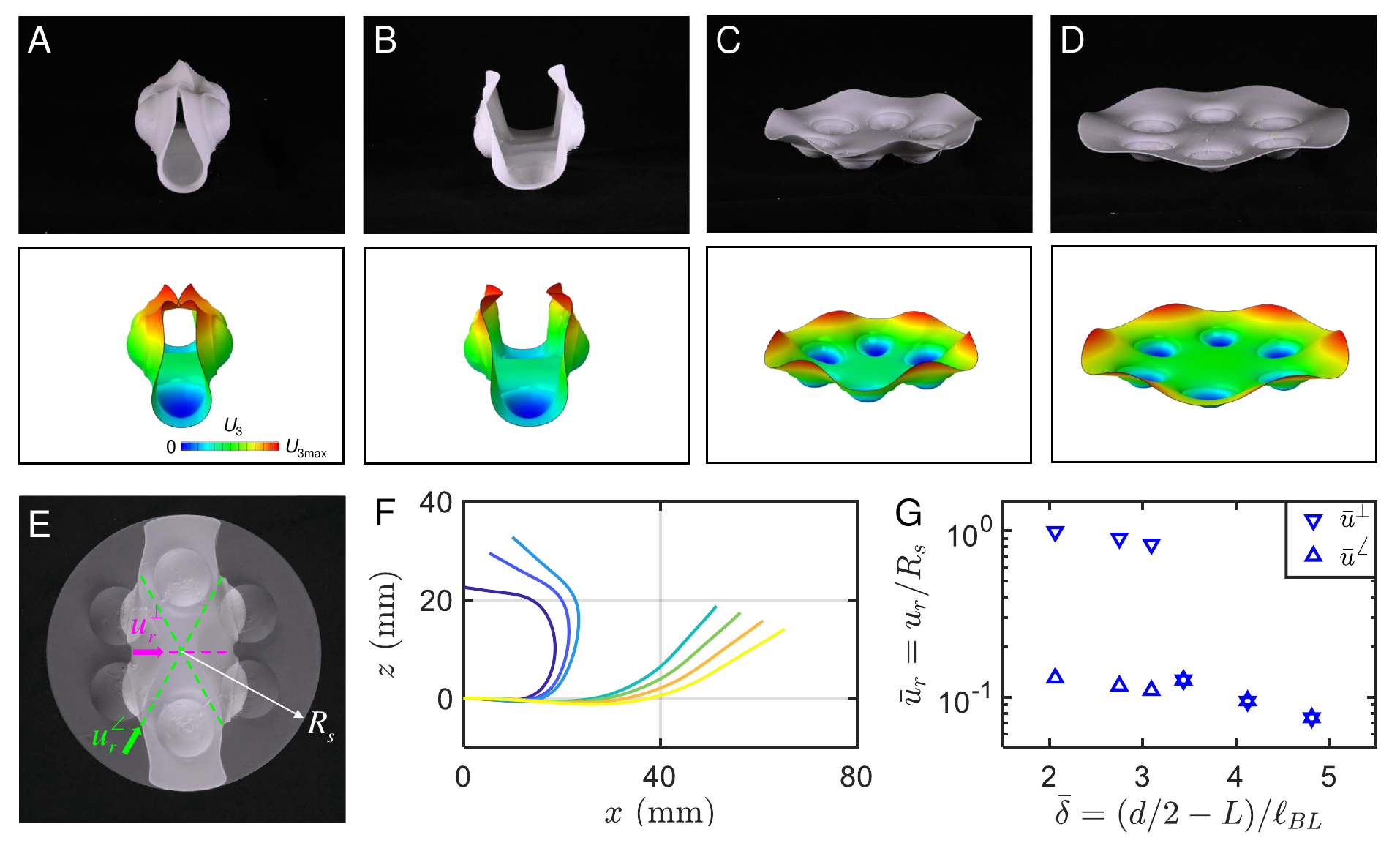}
    \caption{Morphed shape  of soft dimpled sheets with dimples arranged on a hexagonal lattice. (A-D) Experiments (top row) and FEM simulations (bottom row) of snapped dimpled sheets (each with $\alpha = 1.08$, $\lambdad = 8.25$, $L=12\mathrm{~mm}$ fixed) with different dimple spacing; from (A) to (D), $d =30,~33,~34 $ and $40\mathrm{~mm}$, in which, $U_{\rm 3max} = 49.05,~54.66,~31.29,~29.38~{\rm mm}$, respectively. (E) Overlaid plan views of the original (background) and snapped (foreground) shapes of a dimpled sheet with $d = 33\mathrm{~mm}$. The magenta and green dashed lines show the directions in which the horizontal displacements in the directions perpendicular ($\perp$) and inclined ($\angle$)  to the major curved axis are measured. (F) The profiles of snapped sheets taken along the perpendicular axis (i.e.~the magenta line of panel E) with dimple spacing $d$ in the range $30\mathrm{~mm}\leq d\leq40\mathrm{~mm}$. (G) Normalized horizontal displacement of the snapped dimpled sheets   $\bar{u}_r^\perp$ and  $\bar{u}_r^\angle$ (as denoted in panel E), as functions of the normalized dimple spacing, $\bar{\delta}=(d/2-L)/\ell_{BL}$. Note that the cases shown in (A)-(D) correspond to $\bar{\delta}=$ 2.06, 3.09, 3.44 and 5.50, respectively.}
    \label{fig:GlobalMorphingQuantification}
\end{figure}

The cause of the bifurcation from rotational symmetry to the two-fold symmetry discussed above is the high level of strain within the sheet that is caused by the interaction of neighbouring elements. However, we find that, unlike the bilayer sheet, it is not simply the bare strain level that affects the point at which this bifurcation occurs. To demonstrate this, we consider removing a single snapping element from  the centre of the sheet; this  results in a hexagonal lattice as shown in Fig.~\ref{fig:GlobalMorphingQuantification}E. The hexagonal lattice maintains the rotational symmetry of the triangular lattice already considered.

Figs.~ \ref{fig:GlobalMorphingQuantification}A-D demonstrate that the dimpled sheet based on a hexagonal lattice shows behaviour that is phenomenologically similar to that observed for a triangular lattice. In particular, the global shape retains rotational symmetry for large dimple spacing $\bar{\delta}$ but at smaller values of $\bar{\delta}$ a bifurcation to a cylindrical shape occurs (compare Figs.~\ref{fig:GlobalMorphingQuantification}A,B and C,D, for example). The profiles of the snapped dimpled sheets with $d\in[30,40]\mathrm{~mm}$ are also plotted in Fig.~\ref{fig:GlobalMorphingQuantification}F. Comparing to the triangular lattice, the transition from the rotational symmetry to the two-fold symmetry is much sharper for the hexagonal lattice. We also measure the horizontal displacements along two different directions, $u_r^\perp$ and $u_r^\angle$, and plot them as functions of $\bar{\delta}$ in Fig.~\ref{fig:GlobalMorphingQuantification}G. In this case, the bifurcation occurs with $\bar{\delta}\approx3.2$ (compared to $\bar{\delta}\approx6$ for dimples on a triangular lattice). Initially, this smaller separation at bifurcation might seem natural: there are only six dimples in the hexagonal lattice, rather than seven used in the triangular lattice --- since the inversion of the dimples is the root of the strain and hence shape bifurcation, having fewer of them will thus delay the shape bifurcation. However, a more detailed analysis shows that even after accounting for this difference the hexagonal lattice delays the shape bifurcation, as we now show.

\subsection{The effect of lattice spacing}

The standard theory for the buckling of a bilayer sheet \cite[][]{Freund2006} leads to the conclusion that the buckling transition occurs at a fixed strain level. It is therefore natural to compare the strain levels between the cases of triangular and hexagonal lattices. Since the strain level within the inverted dimpled sheet is far from uniform, we seek a measure that gives a value of the strain averaged over the whole sheet and use the total elastic energy, $U_e=\tfrac{1}{2}\int_V\sigma_{ij}\epsilon_{ij}~\mathrm{d}V=\tfrac{1}{2}E\int_V\epsilon_{ij}^2~\mathrm{d}V$, of the sheet (with volume $V$). We therefore introduce a measure of the volume-averaged squared-strain, $\langle\epsilon_{ij}^2\rangle$ defined as
\begin{equation}
    \psi_e=2\langle\epsilon_{ij}^2\rangle=\frac{U_e}{E\cdot V}.
\end{equation}

\begin{figure}[ht]
    \centering
    \vspace{0.5cm}
    \includegraphics[width=0.85\linewidth]{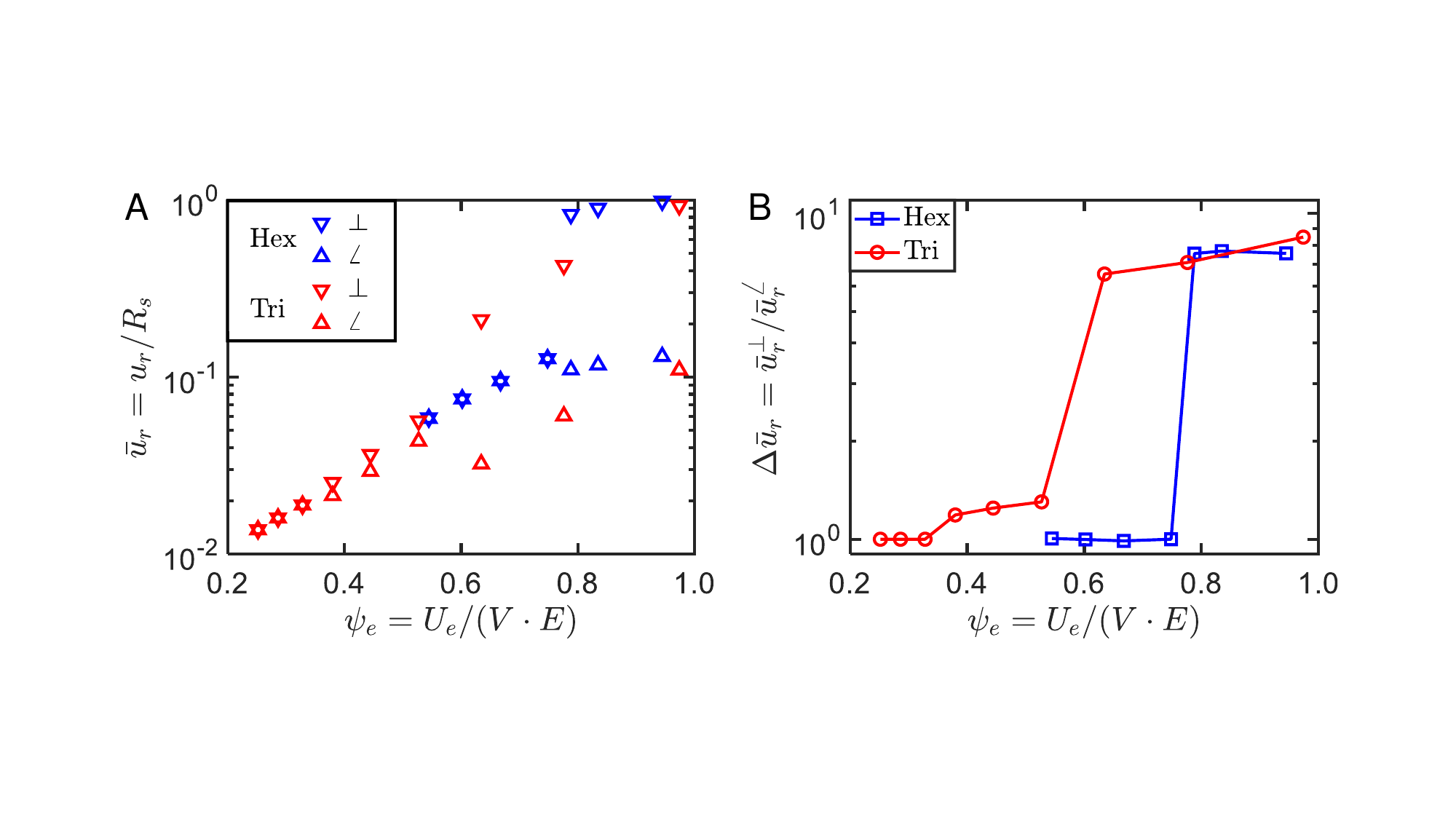}
    \caption{ (A) The normalized radial displacements ($\bar{u}_r$) of the snapped dimpled sheets with both hexagonal and triangular patterns along perpendicular and  inclined directions as a function of the strain energy density ($\psi_e$) obtained from FEM simulations. (B) The ratio between the normalized radial displacements along perpendicular and  inclined directions ($\Delta \bar{u}_r^\perp/\bar{u}_r^\angle$) as a function of the strain energy density ($\psi_e$). Both quantities show a transition, corresponding to global shape bifurcation, at a critical strain energy density; this critical value is higher with a hexagonal lattice than for a triangular lattice.}
    \label{fig:GlobalMorphingComparison}
\end{figure}

Figure \ref{fig:GlobalMorphingComparison} shows how the anisotropy of the deformed shape, as measured by the difference between $u_r^\perp$ and $u_r^\angle$, varies as the volume-averaged strain, encoded by $\psi_e$, varies. This comparison confirms that the shape bifurcation occurs significantly earlier (i.e.~at lower strain) for the triangular lattice than for the hexagonal lattice. Crucially, this measure accounts properly for the fact that there are fewer snapping elements in this case so that the resulting effect can be attributed solely to the geometry of the lattice packing of snapping elements.

\section{Discussion and Conclusions}
\label{sec:DiscussionandConclusion}

In this paper, we have considered the properties of bistable elements (snappits) embedded within a larger soft, shapeable sheet. Such embedded elements are able to induce a global deformation of the resulting dimpled sheet when they switch from their natural state to their other, inverted, state: in this way, `activating' snappits allows them to induce deformation within the sheet.

We began by studying the deformation of a single, activated snappit in the framework of shallow shell theory. In particular, we showed that the lateral scale of the deformation induced by the inversion of a snappit scales with the Pogorelov length scale $\lp=(tR)^{1/2}$. Moreover, by performing systematic FEM simulations, combined with the predictions of shallow shell theory, we found that the bistability of a single snappit is largely governed by the lateral size of the snappit relative to the Pogorelov scale, $\lambda\sim L/\lp$: the size of the outer skirt region has only a relatively weak influence on the threshold value of $\lambda$ at the transition from monostable to bistable (see Fig.~\ref{fig:Bistability}).

Our analysis of the deformation of a single activated snappit showed an azimuthal instability that  breaks the rotational symmetry. Using a boundary layer analysis of the shallow shell equations, we showed that this azimuthal instability results from a large compressive hoop stress, though understanding the mode number of instability remains an open problem. We also showed that when several snappits are combined in a matrix, moderate compressive strains (achieved with moderately spaced snappits) lead to a global buckling mode that respects the rotational symmetry of the underlying lattice; this contrasts with the breaking of rotational symmetry in the single snappit case. Nevertheless, at larger values of the compressive strain (achieved with closely spaced snappits) the sheet deforms globally in a cylindrical shape, breaking the inherent rotational symmetry. Qualitatively, this transition occurs when the boundary layers surrounding each snappit no longer overlap and is  similar to the  shape bifurcation that is well known in a bilayer sheet with mismatch strain \cite[][]{Freund2006}. We also showed that the average level of strain at which the shape transition occurs can be increased by modifying the lattice on which snappits lie: on a hexagonal, rather than triangular, lattice, the strain at which buckling occurs is almost doubled. Controlling  shape bifurcation in shape-shifting structures in this way could open the door to more robust morphing strategies.

\section*{CRediT authorship contribution statement}
\textbf{Mingchao Liu}: Formal analysis, Software, Validation, Writing – original draft. \textbf{Lucie Domino}: Experimental analysis, Methodology, Writing – review \& editing. \textbf{Iris Dupont de Dinechin}: Experimental analysis, Methodology. \textbf{Matteo Taffetani}: Formal analysis, Software. \textbf{Dominic Vella}: Conceptualization, Methodology, Supervision, Writing – review \& editing.

\section*{Declaration of Competing Interest}
The authors declare that they have no known competing financial interests or personal relationships that could have appeared to influence the work reported in this paper.

\section*{Data availability}
Data will be made available on request.

\section*{Acknowledgments}
The research leading to these results has received funding from the Royal Society through a Newton International Fellowship (ML), the European Research Council under the European Union's Horizon 2020 Programme / ERC Grant Agreement no.~637334 (DV) and the Leverhulme Trust through a Philip Leverhulme Prize (DV). MT is a member of the Gruppo Nazionale di Fisica Matematica (GNFM) of the Istituto Nazionale di Alta Matematica (INdAM). We are grateful to Y.~Zhang for discussions on FEA simulations, and to S. Koot for his help on 3D printing samples.

\begin{appendix}
\setcounter{figure}{0}
\setcounter{equation}{0}
\setcounter{table}{0}
\renewcommand{\thefigure}{A.\arabic{figure}}
\renewcommand{\theequation}{A.\arabic{equation}}

\section{Experimental Details}
\label{methods_exp}

The silicone mould shown in figure \ref{fig:ChocolateMould} is a commercially available chocolate mould. In all other experiments described in this paper, physical samples were 3D printed using a flexible thermoplastic polymer (Filaflex 60A) using desktop 3D printers (Wanhao Duplicator i3 and Tool Changer --- E3D). The reference and deformed shapes were measured with a custom 3D scanning technique based on Fourier Transform Profilometry, a structured light method that uses stripes shone on an object to measure its 3D shape \cite[][]{Jeught2016}.

\section{FEM Simulations}
\label{methods_fem}
Computational analyses were developed using the commercial finite element analysis software (ABAQUS) to simulate the deformation of both a snappit (i.e. a single plate/shell element) and a dimpled sheet (consisting of many snappits). The type of element used to represent the system depended on the problem under consideration (see below). Typically CAX4RH and S4R elements were used for axisymmetric  and 3D shell models, respectively. Convergence was tested (by comparison with refined meshes) to ensure computational accuracy. The problem is essentially geometrical since no external forces were imposed. Within FEM material constants of density $\rho = 1.3 \times 10^3\mathrm{~kg/m^3}$, Young's modulus $E = 6\mathrm{~MPa}$ and Poisson's ratio $\nu = 0.4999$ were taken for the majority of simulations, which were performed using the linearly elastic constitutive relation. To assess the importance of material nonlinearity effects, some simulations were also performed using the Neo-Hookean constitutive model with $C_{10} = 1\mathrm{~MPa}$ and $D_1 = 0$. 

\subsection{Inversion of a single snappit}
To model the axisymmetric deformations of a single snappit (Section II), we used 4-node hybrid bilinear axisymmetric quadrilateral elements (CAX4RH) for both the shell and the plate. Simulations were conducted in two steps (Static, General): first we applied pressure load to the top surface of the shell to invert the shell. The pressure was then removed in the second step, allowing the system to relax and find the closest equilibrium solution (an artificial global damping factor of $\etastab= 1 \times 10^{-8}$ was specified to stabilize the simulations). We consider two types of boundary conditions at the outer edge: (i) a hinged boundary, i.e.~the vertical displacement and bending moment vanish ($u_z = M_r = 0$) at the outer edge ($r = L + \Delta L$) and, (ii) a clamped boundary, i.e.~both the vertical displacement and the radial rotation vanish ($u_z = \partial u_{z}/\partial r = 0$) at the outer edge. For some combinations of geometrical parameters, the system returns to its initial, undeformed state after relaxation. In this case, the system is monostable (since no alternative equilibrium to the natural undeformed state has been found). If an alternative equilibrium is found in this way, the system is bistable. In the bistable scenario, the inverted shape is extracted from the bottom surface of the shape; this is consistent with what is measured experimentally.

\subsection{Buckling of a single snappit and of a dimpled sheet}
For the azuimuthal (non-axisymmetric) buckling of the snappit (in Section \ref{sec:Origin_instability}) and the global deformation of the dimpled sheet induced by the buckling of each dimples (in Section \ref{sec:ManyElements}), we used 4-node doubly curved shell elements (S4R). The simulations were also conducted in three steps (here we chose the procedure type of Dynamic, Implicit): for a single snappit, we fixed the outer edge ($r = L + \Delta L$) and applied a displacement load at the centre ($r = 0$) in the first step, where the displacement is prescribed as twice the initial height of the shell (i.e. $u_z ~= W = 2(R-H)$); we then removed the displacement load and released the fixed boundary condition at the outer edge in the second step, fixing the centre at the position as prescribed in the first step, allowing the system to relax; then a third step for relaxing the system is set to make sure the closest equilibrium solution is reached. For a dimpled sheet, the outer edge and central point were fixed, and a displacement load  applied at the centre of each dimple (again, $u_z = 2(R-H)$) in the first step; in the second step, we removed the displacement and released the fixed boundary condition at the outer edge to let the system  relax and find the closest equilibrium solution. We then extracted the profile of the curved shape of the sheet in the snapped state, and measured the radius of curvature by fitting the extracted profiles.

\section{Boundary conditions\label{sec:AppB3}}

To make the problem solved numerically in Section \ref{sec:SingleElement} clearer, we summarize the appropriate boundary conditions, expressed in
terms of the unknowns of the problem, together with their physical meaning in Table \ref{tab:Boundarycondition}.

\begin{table}[h]

    \centering
\caption{Boundary conditions required for solving the problem in the shallow shell formulation of Section \ref{sec:SingleElement}.}
	\label{tab:Boundarycondition}
	\setlength{\tabcolsep}{5pt}
	\begin{tabular}{|l|l|l|}
	\hline
	Position & Condition & Description\\
	\hline
    \multirow{3}{1.7em}{$\rho = 0$} & $W(0)=0$ & Zero vertical displacement\\
    \cline{2-3}
    & $W'(0)= \frac{\upd W}{\upd \rho}\big|_{\rho = 0}= 0$ & Zero slope\\
    \cline{2-3}
    & $U_r(0) = \rho \frac{\upd \Psi}{\upd \rho} - \nu \Psi \big|_{\rho = 0} = 0$ & Zero radial displacement\\
	\hline
	\multirow{6}{1.7em}{$\rho = 1$} & $\left[W \right]_{1+}^{1-} = 0$ & Continuous vertical displacement\\
    \cline{2-3}
    & $\left[U_r \right]_{1+}^{1-} = \left[\rho \frac{\upd \Psi}{\upd \rho} - \nu \Psi \right]_{1+}^{1-} = 0$ & Continuous horizontal displacement\\
    \cline{2-3}
    & $\left[\sigma_{rr} \right]_{1+}^{1-} = \left[\frac{1}{\rho} \Psi \right]_{1+}^{1-}=0 $ & Continuous radial stress\\
    \cline{2-3}
    & $\left[M\right]_{1+}^{1-} = \left[\frac{\upd^2 W}{\upd\rho^2} + \frac{\nu}{\rho} \frac{\upd W}{\upd\rho}\right]_{1+}^{1-}=0$ & Continuous bending moment\\
    \cline{2-3}
    & $\left[F_r \right]_{1+}^{1-} = \left[\frac{\upd^3 W}{\upd\rho^3}+\frac{1}{\rho} \frac{\upd^2 W}{\upd\rho^2} - \frac{1}{\rho^2} \frac{\upd W}{\upd\rho} \right]_{1+}^{1-} = 0$ & Continuous shear force\\
    \cline{2-3}
    & $\left[W' \right]_{1+}^{1-} = \left[\frac{\upd W}{\upd \rho}\right]_{1+}^{1-} = 0$ & Continuous slope of the displacement\\
    \hline
    \multirow{3}{1.7em}{$\rho = 1+\Delta$} & $M(1+\Delta)=0$ & Zero bending moment\\
    \cline{2-3}
    & $F_r(1+\Delta) = \frac{\upd^3 W}{\upd\rho^3}+\frac{1}{\rho} \frac{\upd^2 W}{\upd\rho^2} - \frac{1}{\rho^2} \frac{\upd W}{\upd\rho}\big|_{\rho = 1+\Delta} = 0$ & Zero shear force\\
    \cline{2-3}
    & $\sigma_{rr}(1+\Delta) = \frac{1}{\rho} \Psi\big|_{\rho = 1+\Delta} = 0$ & Zero radial stress\\
    \hline
    \end{tabular}
\end{table}

\end{appendix}

\bibliographystyle{elsarticle-num-names}
\bibliography{Bibliography.bib}

\begin{thebibliography}{44}
\expandafter\ifx\csname natexlab\endcsname\relax\def\natexlab#1{#1}\fi
\providecommand{\url}[1]{\texttt{#1}}
\providecommand{\href}[2]{#2}
\providecommand{\path}[1]{#1}
\providecommand{\DOIprefix}{doi:}
\providecommand{\ArXivprefix}{arXiv:}
\providecommand{\URLprefix}{URL: }
\providecommand{\Pubmedprefix}{pmid:}
\providecommand{\doi}[1]{\href{http://dx.doi.org/#1}{\path{#1}}}
\providecommand{\Pubmed}[1]{\href{pmid:#1}{\path{#1}}}
\providecommand{\bibinfo}[2]{#2}
\ifx\xfnm\relax \def\xfnm[#1]{\unskip,\space#1}\fi
%Type = Article
\bibitem[{Alapan et~al.(2020)Alapan, Karacakol, Guzelhan, Isik, and
  Sitti}]{Alapan2020}
\bibinfo{author}{Y.~Alapan}, \bibinfo{author}{A.~C. Karacakol},
  \bibinfo{author}{S.~N. Guzelhan}, \bibinfo{author}{I.~Isik},
  \bibinfo{author}{M.~Sitti},
\newblock \bibinfo{title}{Reprogrammable shape morphing of magnetic soft
  machines},
\newblock \bibinfo{journal}{Sci. Adv.} \bibinfo{volume}{6}
  (\bibinfo{year}{2020}) \bibinfo{pages}{eabc6414}.
%Type = Article
\bibitem[{Shah et~al.(2021)Shah, Powers, Tilton, Kriegman, Bongard, and
  Kramer-Bottiglio}]{Shah2021}
\bibinfo{author}{D.~S. Shah}, \bibinfo{author}{J.~P. Powers},
  \bibinfo{author}{L.~G. Tilton}, \bibinfo{author}{S.~Kriegman},
  \bibinfo{author}{J.~Bongard}, \bibinfo{author}{R.~Kramer-Bottiglio},
\newblock \bibinfo{title}{A soft robot that adapts to environments through
  shape change},
\newblock \bibinfo{journal}{Nat. Mach. Intell.} \bibinfo{volume}{3}
  (\bibinfo{year}{2021}) \bibinfo{pages}{51--59}.
%Type = Article
\bibitem[{Liu et~al.(2021)Liu, Hacker, and Daraio}]{liu2021}
\bibinfo{author}{K.~Liu}, \bibinfo{author}{F.~Hacker},
  \bibinfo{author}{C.~Daraio},
\newblock \bibinfo{title}{Robotic surfaces with reversible, spatiotemporal
  control for shape morphing and object manipulation},
\newblock \bibinfo{journal}{Sci. Robot.} \bibinfo{volume}{6}
  (\bibinfo{year}{2021}) \bibinfo{pages}{eabf5116}.
%Type = Article
\bibitem[{Pikul et~al.(2017)Pikul, Li, Bai, Hanlon, Cohen, and
  Shepherd}]{Pikul2017}
\bibinfo{author}{J.~Pikul}, \bibinfo{author}{S.~Li}, \bibinfo{author}{H.~Bai},
  \bibinfo{author}{R.~Hanlon}, \bibinfo{author}{I.~Cohen},
  \bibinfo{author}{R.~Shepherd},
\newblock \bibinfo{title}{Stretchable surfaces with programmable 3d texture
  morphing for synthetic camouflaging skins},
\newblock \bibinfo{journal}{Science} \bibinfo{volume}{358}
  (\bibinfo{year}{2017}) \bibinfo{pages}{210--214}.
%Type = Article
\bibitem[{Si{\'e}fert et~al.(2019)Si{\'e}fert, Reyssat, Bico, and
  Roman}]{siefert2019}
\bibinfo{author}{E.~Si{\'e}fert}, \bibinfo{author}{E.~Reyssat},
  \bibinfo{author}{J.~Bico}, \bibinfo{author}{B.~Roman},
\newblock \bibinfo{title}{Bio-inspired pneumatic shape-morphing elastomers},
\newblock \bibinfo{journal}{Nat. Mater.} \bibinfo{volume}{18}
  (\bibinfo{year}{2019}) \bibinfo{pages}{24--28}.
%Type = Article
\bibitem[{Boley et~al.(2019)Boley, van Rees, Lissandrello, Horenstein, Truby,
  Kotikian, Lewis, and Mahadevan}]{boley2019}
\bibinfo{author}{J.~W. Boley}, \bibinfo{author}{W.~M. van Rees},
  \bibinfo{author}{C.~Lissandrello}, \bibinfo{author}{M.~N. Horenstein},
  \bibinfo{author}{R.~L. Truby}, \bibinfo{author}{A.~Kotikian},
  \bibinfo{author}{J.~A. Lewis}, \bibinfo{author}{L.~Mahadevan},
\newblock \bibinfo{title}{Shape-shifting structured lattices via multimaterial
  4d printing},
\newblock \bibinfo{journal}{Proc. Natl Acad. Sci. USA} \bibinfo{volume}{116}
  (\bibinfo{year}{2019}) \bibinfo{pages}{20856--20862}.
%Type = Article
\bibitem[{Zhang et~al.(2021)Zhang, Guo, Hu, and Sitti}]{zhang2021}
\bibinfo{author}{J.~Zhang}, \bibinfo{author}{Y.~Guo}, \bibinfo{author}{W.~Hu},
  \bibinfo{author}{M.~Sitti},
\newblock \bibinfo{title}{Wirelessly actuated thermo-and magneto-responsive
  soft bimorph materials with programmable shape-morphing},
\newblock \bibinfo{journal}{Adv. Mater.} \bibinfo{volume}{33}
  (\bibinfo{year}{2021}) \bibinfo{pages}{2100336}.
%Type = Article
\bibitem[{Kim et~al.(2018)Kim, Yuk, Zhao, Chester, and Zhao}]{kim2018}
\bibinfo{author}{Y.~Kim}, \bibinfo{author}{H.~Yuk}, \bibinfo{author}{R.~Zhao},
  \bibinfo{author}{S.~A. Chester}, \bibinfo{author}{X.~Zhao},
\newblock \bibinfo{title}{Printing ferromagnetic domains for untethered
  fast-transforming soft materials},
\newblock \bibinfo{journal}{Nature} \bibinfo{volume}{558}
  (\bibinfo{year}{2018}) \bibinfo{pages}{274--279}.
%Type = Article
\bibitem[{Aharoni et~al.(2018)Aharoni, Xia, Zhang, Kamien, and
  Yang}]{aharoni2018}
\bibinfo{author}{H.~Aharoni}, \bibinfo{author}{Y.~Xia},
  \bibinfo{author}{X.~Zhang}, \bibinfo{author}{R.~D. Kamien},
  \bibinfo{author}{S.~Yang},
\newblock \bibinfo{title}{Universal inverse design of surfaces with thin
  nematic elastomer sheets},
\newblock \bibinfo{journal}{Proc. Natl Acad. Sci. USA} \bibinfo{volume}{115}
  (\bibinfo{year}{2018}) \bibinfo{pages}{7206--7211}.
%Type = Article
\bibitem[{Celli et~al.(2018)Celli, McMahan, Ramirez, Bauhofer, Naify, Hofmann,
  Audoly, and Daraio}]{celli2018}
\bibinfo{author}{P.~Celli}, \bibinfo{author}{C.~McMahan},
  \bibinfo{author}{B.~Ramirez}, \bibinfo{author}{A.~Bauhofer},
  \bibinfo{author}{C.~Naify}, \bibinfo{author}{D.~Hofmann},
  \bibinfo{author}{B.~Audoly}, \bibinfo{author}{C.~Daraio},
\newblock \bibinfo{title}{Shape-morphing architected sheets with non-periodic
  cut patterns},
\newblock \bibinfo{journal}{Soft Matter} \bibinfo{volume}{14}
  (\bibinfo{year}{2018}) \bibinfo{pages}{9744--9749}.
%Type = Article
\bibitem[{Liu et~al.(2020)Liu, Domino, and Vella}]{liu2020}
\bibinfo{author}{M.~Liu}, \bibinfo{author}{L.~Domino},
  \bibinfo{author}{D.~Vella},
\newblock \bibinfo{title}{Tapered elastic{\ae} as a route for axisymmetric
  morphing structures},
\newblock \bibinfo{journal}{Soft Matter} \bibinfo{volume}{16}
  (\bibinfo{year}{2020}) \bibinfo{pages}{7739--7750}.
%Type = Article
\bibitem[{Wang et~al.(2017)Wang, Zhu, Hong, Wu, and Zheng}]{wang2017}
\bibinfo{author}{Z.~J. Wang}, \bibinfo{author}{C.~N. Zhu},
  \bibinfo{author}{W.~Hong}, \bibinfo{author}{Z.~L. Wu},
  \bibinfo{author}{Q.~Zheng},
\newblock \bibinfo{title}{Cooperative deformations of periodically patterned
  hydrogels},
\newblock \bibinfo{journal}{Science Advances} \bibinfo{volume}{3}
  (\bibinfo{year}{2017}) \bibinfo{pages}{e1700348}.
%Type = Article
\bibitem[{Faber et~al.(2020)Faber, Udani, Riley, Studart, and
  Arrieta}]{faber2020}
\bibinfo{author}{J.~A. Faber}, \bibinfo{author}{J.~P. Udani},
  \bibinfo{author}{K.~S. Riley}, \bibinfo{author}{A.~R. Studart},
  \bibinfo{author}{A.~F. Arrieta},
\newblock \bibinfo{title}{Dome-patterned metamaterial sheets},
\newblock \bibinfo{journal}{Adv. Sci.} \bibinfo{volume}{7}
  (\bibinfo{year}{2020}) \bibinfo{pages}{2001955}.
%Type = Article
\bibitem[{Moessner and Ramirez(2006)}]{moessner2006}
\bibinfo{author}{R.~Moessner}, \bibinfo{author}{A.~P. Ramirez},
\newblock \bibinfo{title}{Geometrical frustration},
\newblock \bibinfo{journal}{Phys. Today} \bibinfo{volume}{59}
  (\bibinfo{year}{2006}) \bibinfo{pages}{24}.
%Type = Article
\bibitem[{Gilbert et~al.(2016)Gilbert, Nisoli, and Schiffer}]{gilbert2016}
\bibinfo{author}{I.~Gilbert}, \bibinfo{author}{C.~Nisoli},
  \bibinfo{author}{P.~Schiffer},
\newblock \bibinfo{title}{Frustration by design},
\newblock \bibinfo{journal}{Phys. Today} \bibinfo{volume}{69}
  (\bibinfo{year}{2016}) \bibinfo{pages}{54}.
%Type = Article
\bibitem[{Si{\'e}fert et~al.(2021)Si{\'e}fert, Levin, and Sharon}]{siefert2021}
\bibinfo{author}{E.~Si{\'e}fert}, \bibinfo{author}{I.~Levin},
  \bibinfo{author}{E.~Sharon},
\newblock \bibinfo{title}{Euclidean frustrated ribbons},
\newblock \bibinfo{journal}{Phys. Rev. X} \bibinfo{volume}{11}
  (\bibinfo{year}{2021}) \bibinfo{pages}{011062}.
%Type = Article
\bibitem[{Holmes and Crosby(2007)}]{holmes2007}
\bibinfo{author}{D.~P. Holmes}, \bibinfo{author}{A.~J. Crosby},
\newblock \bibinfo{title}{Snapping surfaces},
\newblock \bibinfo{journal}{Adv. Mater.} \bibinfo{volume}{19}
  (\bibinfo{year}{2007}) \bibinfo{pages}{3589--3593}.
%Type = Article
\bibitem[{Taffetani et~al.(2018)Taffetani, Jiang, Holmes, and
  Vella}]{Taffetani2018}
\bibinfo{author}{M.~Taffetani}, \bibinfo{author}{X.~Jiang},
  \bibinfo{author}{D.~P. Holmes}, \bibinfo{author}{D.~Vella},
\newblock \bibinfo{title}{Static bistability of spherical caps},
\newblock \bibinfo{journal}{Proc. R. Soc. A} \bibinfo{volume}{474}
  (\bibinfo{year}{2018}) \bibinfo{pages}{20170910}.
%Type = Article
\bibitem[{Liu et~al.(2021)Liu, Gomez, and Vella}]{liu2021delayed}
\bibinfo{author}{M.~Liu}, \bibinfo{author}{M.~Gomez},
  \bibinfo{author}{D.~Vella},
\newblock \bibinfo{title}{Delayed bifurcation in elastic snap-through
  instabilities},
\newblock \bibinfo{journal}{J Mech. Phys. Solids} \bibinfo{volume}{151}
  (\bibinfo{year}{2021}) \bibinfo{pages}{104386}.
%Type = Article
\bibitem[{Oshri et~al.(2019)Oshri, Biswas, and Balazs}]{Oshri2019}
\bibinfo{author}{O.~Oshri}, \bibinfo{author}{S.~Biswas}, \bibinfo{author}{A.~C.
  Balazs},
\newblock \bibinfo{title}{Modeling the behavior of inclusions in circular
  plates undergoing shape changes from two to three dimensions},
\newblock \bibinfo{journal}{Phys. Rev. E} \bibinfo{volume}{100}
  (\bibinfo{year}{2019}) \bibinfo{pages}{043001}.
%Type = Article
\bibitem[{Plummer and Nelson(2020)}]{Plummer2020}
\bibinfo{author}{A.~Plummer}, \bibinfo{author}{D.~R. Nelson},
\newblock \bibinfo{title}{Buckling and metastability in membranes with dilation
  arrays},
\newblock \bibinfo{journal}{Phys. Rev. E} \bibinfo{volume}{102}
  (\bibinfo{year}{2020}) \bibinfo{pages}{033002}.
%Type = Article
\bibitem[{Oshri et~al.(2020)Oshri, Biswas, and Balazs}]{Oshri2020}
\bibinfo{author}{O.~Oshri}, \bibinfo{author}{S.~Biswas}, \bibinfo{author}{A.~C.
  Balazs},
\newblock \bibinfo{title}{Buckling-induced interaction between circular
  inclusions in an infinite thin plate},
\newblock \bibinfo{journal}{Phys. Rev. E} \bibinfo{volume}{102}
  (\bibinfo{year}{2020}) \bibinfo{pages}{033004}.
%Type = Article
\bibitem[{Hanakata et~al.(2022)Hanakata, Plummer, and Nelson}]{hanakata2022}
\bibinfo{author}{P.~Z. Hanakata}, \bibinfo{author}{A.~Plummer},
  \bibinfo{author}{D.~R. Nelson},
\newblock \bibinfo{title}{Anomalous thermal expansion in ising-like puckered
  sheets},
\newblock \bibinfo{journal}{Physical Review Letters} \bibinfo{volume}{128}
  (\bibinfo{year}{2022}) \bibinfo{pages}{075902}.
%Type = Article
\bibitem[{Seffen(2006)}]{seffen2006}
\bibinfo{author}{K.~Seffen},
\newblock \bibinfo{title}{Mechanical memory metal: a novel material for
  developing morphing engineering structures},
\newblock \bibinfo{journal}{Scripta Mater.} \bibinfo{volume}{55}
  (\bibinfo{year}{2006}) \bibinfo{pages}{411--414}.
%Type = Article
\bibitem[{Seffen(2007)}]{seffen2007}
\bibinfo{author}{K.~Seffen},
\newblock \bibinfo{title}{Hierarchical multi-stable shapes in mechanical memory
  metal},
\newblock \bibinfo{journal}{Scripta Mater.} \bibinfo{volume}{56}
  (\bibinfo{year}{2007}) \bibinfo{pages}{417--420}.
%Type = Article
\bibitem[{Gorissen et~al.(2020)Gorissen, Melancon, Vasios, Torbati, and
  Bertoldi}]{gorissen2020}
\bibinfo{author}{B.~Gorissen}, \bibinfo{author}{D.~Melancon},
  \bibinfo{author}{N.~Vasios}, \bibinfo{author}{M.~Torbati},
  \bibinfo{author}{K.~Bertoldi},
\newblock \bibinfo{title}{Inflatable soft jumper inspired by shell snapping},
\newblock \bibinfo{journal}{Sci. Robot.} \bibinfo{volume}{5}
  (\bibinfo{year}{2020}) \bibinfo{pages}{eabb1967}.
%Type = Article
\bibitem[{Chen et~al.(2021)Chen, Pauly, and Reis}]{Chen2021}
\bibinfo{author}{T.~Chen}, \bibinfo{author}{M.~Pauly}, \bibinfo{author}{P.~M.
  Reis},
\newblock \bibinfo{title}{A reprogrammable mechanical metamaterial with stable
  memory},
\newblock \bibinfo{journal}{Nature} \bibinfo{volume}{589}
  (\bibinfo{year}{2021}) \bibinfo{pages}{386---390}.
%Type = Article
\bibitem[{Udani and Arrieta(2022)}]{udani2022taming}
\bibinfo{author}{J.~P. Udani}, \bibinfo{author}{A.~F. Arrieta},
\newblock \bibinfo{title}{Taming geometric frustration by leveraging structural
  elasticity},
\newblock \bibinfo{journal}{Materials \& Design} \bibinfo{volume}{221}
  (\bibinfo{year}{2022}) \bibinfo{pages}{110809}.
%Type = Book
\bibitem[{Ventsel and Krauthammer(2001)}]{ventsel2001}
\bibinfo{author}{E.~Ventsel}, \bibinfo{author}{T.~Krauthammer},
  \bibinfo{title}{Thin plates and shells}, \bibinfo{publisher}{NY: Marcel
  Dekker}, \bibinfo{year}{2001}.
%Type = Book
\bibitem[{Libai and Simmonds(2005)}]{libai2005}
\bibinfo{author}{A.~Libai}, \bibinfo{author}{J.~G. Simmonds},
  \bibinfo{title}{The nonlinear theory of elastic shells},
  \bibinfo{publisher}{Cambridge university press}, \bibinfo{year}{2005}.
%Type = Phdthesis
\bibitem[{Sobota(2020)}]{sobota2020}
\bibinfo{author}{P.~Sobota}, \bibinfo{title}{Multistable shell structures},
  Ph.D. thesis, University of Cambridge, \bibinfo{year}{2020}.
%Type = Book
\bibitem[{Calladine(1989)}]{calladine1989}
\bibinfo{author}{C.~R. Calladine}, \bibinfo{title}{Theory of shell structures},
  \bibinfo{publisher}{Cambridge university press}, \bibinfo{year}{1989}.
%Type = Techreport
\bibitem[{Doedel et~al.(2007)Doedel, Champneys, Dercole, Fairgrieve, Kuznetsov,
  Oldeman, Paffenroth, Sandstede, Wang, and Zhang}]{doedel2007}
\bibinfo{author}{E.~J. Doedel}, \bibinfo{author}{A.~R. Champneys},
  \bibinfo{author}{F.~Dercole}, \bibinfo{author}{T.~F. Fairgrieve},
  \bibinfo{author}{Y.~A. Kuznetsov}, \bibinfo{author}{B.~Oldeman},
  \bibinfo{author}{R.~Paffenroth}, \bibinfo{author}{B.~Sandstede},
  \bibinfo{author}{X.~Wang}, \bibinfo{author}{C.~Zhang},
  \bibinfo{title}{AUTO-07P: Continuation and bifurcation software for ordinary
  differential equations}, \bibinfo{type}{Technical Report},
  \bibinfo{address}{Montreal, Canada}, \bibinfo{year}{2007}.
%Type = Book
\bibitem[{Pogorelov(1988)}]{Pogorelov1988}
\bibinfo{author}{A.~V. Pogorelov}, \bibinfo{title}{Bendings of surfaces and
  stability of shells}, \bibinfo{publisher}{American Mathematical Soc.},
  \bibinfo{year}{1988}.
%Type = Article
\bibitem[{Gomez et~al.(2016)Gomez, Moulton, and Vella}]{Gomez2016}
\bibinfo{author}{M.~Gomez}, \bibinfo{author}{D.~E. Moulton},
  \bibinfo{author}{D.~Vella},
\newblock \bibinfo{title}{The shallow shell approach to {P}ogorelov's problem
  and the breakdown of `mirror buckling'},
\newblock \bibinfo{journal}{Proc. R. Soc. A} \bibinfo{volume}{472}
  (\bibinfo{year}{2016}) \bibinfo{pages}{20150732}.
%Type = Article
\bibitem[{Coman and Bassom(2007)}]{Coman2007}
\bibinfo{author}{C.~D. Coman}, \bibinfo{author}{A.~P. Bassom},
\newblock \bibinfo{title}{On the wrinkling of a pre-stressed annular thin film
  in tension},
\newblock \bibinfo{journal}{J. Mech. Phys. Solids} \bibinfo{volume}{55}
  (\bibinfo{year}{2007}) \bibinfo{pages}{1601--1617}.
%Type = Article
\bibitem[{Davidovitch et~al.(2011)Davidovitch, Schroll, Vella, Adda-Bedia, and
  Cerda}]{Davidovitch2011}
\bibinfo{author}{B.~Davidovitch}, \bibinfo{author}{R.~D. Schroll},
  \bibinfo{author}{D.~Vella}, \bibinfo{author}{M.~Adda-Bedia},
  \bibinfo{author}{E.~Cerda},
\newblock \bibinfo{title}{Prototypical model for tensional wrinkling in thin
  sheets},
\newblock \bibinfo{journal}{Proc. Natl. Acad. Sci. USA} \bibinfo{volume}{108}
  (\bibinfo{year}{2011}) \bibinfo{pages}{18227--18232}.
%Type = Article
\bibitem[{Box et~al.(2020)Box, Kodio, O'Kiely, Cantelli, Goriely, and
  Vella}]{Box2020}
\bibinfo{author}{F.~Box}, \bibinfo{author}{O.~Kodio},
  \bibinfo{author}{D.~O'Kiely}, \bibinfo{author}{V.~Cantelli},
  \bibinfo{author}{A.~Goriely}, \bibinfo{author}{D.~Vella},
\newblock \bibinfo{title}{Dynamic buckling of an elastic ring in a soap film},
\newblock \bibinfo{journal}{Phys. Rev. Lett.} \bibinfo{volume}{124}
  (\bibinfo{year}{2020}) \bibinfo{pages}{198003}.
%Type = Article
\bibitem[{Pal et~al.(2022)Pal, Pocivavsek, and Witten}]{Pal2022}
\bibinfo{author}{A.~S. Pal}, \bibinfo{author}{L.~Pocivavsek},
  \bibinfo{author}{T.~A. Witten},
\newblock \bibinfo{title}{Faceted wrinkling by contracting a curved boundary},
\newblock \bibinfo{journal}{arXiv preprint, arXiv:2206.03552}
  (\bibinfo{year}{2022}).
%Type = Article
\bibitem[{Klein et~al.(2007)Klein, Efrati, and Sharon}]{Klein2007}
\bibinfo{author}{Y.~Klein}, \bibinfo{author}{E.~Efrati},
  \bibinfo{author}{E.~Sharon},
\newblock \bibinfo{title}{Shaping of elastic sheets by prescription of
  non-{E}uclidean metrics},
\newblock \bibinfo{journal}{{Science}} \bibinfo{volume}{315}
  (\bibinfo{year}{2007}) \bibinfo{pages}{1116--1120}.
%Type = Article
\bibitem[{Dervaux and {Ben Amar}(2008)}]{Dervaux2008}
\bibinfo{author}{J.~Dervaux}, \bibinfo{author}{M.~{Ben Amar}},
\newblock \bibinfo{title}{Morphogenesis of growing soft tissues},
\newblock \bibinfo{journal}{Phys. Rev. Lett.} \bibinfo{volume}{101}
  (\bibinfo{year}{2008}) \bibinfo{pages}{068101}.
%Type = Article
\bibitem[{Sharon and Efrati(2010)}]{Sharon2010}
\bibinfo{author}{E.~Sharon}, \bibinfo{author}{E.~Efrati},
\newblock \bibinfo{title}{The mechanics of non-euclidean plates},
\newblock \bibinfo{journal}{Soft Matter} \bibinfo{volume}{6}
  (\bibinfo{year}{2010}) \bibinfo{pages}{5693–--5704}.
%Type = Book
\bibitem[{Freund and Suresh(2006)}]{Freund2006}
\bibinfo{author}{L.~B. Freund}, \bibinfo{author}{S.~Suresh},
  \bibinfo{title}{Thin Film Materials}, \bibinfo{publisher}{Cambridge
  University Press}, \bibinfo{year}{2006}.
%Type = Article
\bibitem[{Van~der Jeught and Dirckx(2016)}]{Jeught2016}
\bibinfo{author}{S.~Van~der Jeught}, \bibinfo{author}{J.~J. Dirckx},
\newblock \bibinfo{title}{Real-time structured light profilometry: a review},
\newblock \bibinfo{journal}{Optics and Lasers in Engineering}
  \bibinfo{volume}{87} (\bibinfo{year}{2016}) \bibinfo{pages}{18--31}.

\end{thebibliography}

\end{document}